\documentclass[structabstract]{aa}  
\usepackage{graphicx}
\usepackage{txfonts}
\usepackage{natbib}
\usepackage{longtable}

\begin{document}

\title{On the effects of rotation in primordial star-forming clouds}

\author{Jayanta Dutta\inst{1,2}} 

\institute{$^1$Instituto de Astrofísica e Ciências do Espaço, Universidade de Lisboa, OAL, Tapada da Ajuda, PT1349-018 Lisboa, Portugal\\
$^2$Inter-University Centre for Astronomy and Astrophysics, Post Bag 4, 
Ganeshkhind, Pune University Campus, Pune 411007, India\\
\email{jd.astrop@gmail.com}}


\abstract
{The thermodynamical evolution of gas during the collapse of the primordial 
star-forming cloud depends significantly on the initial degree of rotation.}
{However, there is no clear understanding of how the initial rotation can 
affect the heating and cooling process and hence the temperature that leads 
to the fragmentation of the gas during Population III star formation.} 
{We report the results from three\hbox{-}dimensional, smoothed-particle 
hydrodynamics (SPH) simulations of a rotating self-gravitating primordial 
gas cloud with a modified version of the Gadget-2 code, in which the initial 
ratio of the rotational to the gravitational energy ($\beta_0$) is varied 
over two orders of magnitude.}
{We find that despite the lack of any initial turbulence and magnetic 
fields in the clouds, the angular momentum distribution leads to the 
formation and build-up of a disk that fragments into several clumps. 
We further examine the behavior of the protostars that form in both 
idealized as well as more realistic minihalos from the cosmological 
simulations. The thermodynamical evolution and the 
fragmentation behavior of the cosmological minihalos are similar to 
that of the artificial cases, especially in those with a similar 
$\beta_0$-parameter. Protostars with a higher rotation support exhibit 
spiral-arm-like structures on several scales, and have lower accretion 
rates. These type of clouds tend to fragment more, while some of the 
protostars escape from the cluster with the possibility of surviving until 
the present day. They also take much longer to form compared to their 
slowly rotating counterparts.}
{We conclude that the use of appropriate initial conditions of the gas in
minihalos is a pivotal and decisive quantity to study the evolution and
final fate of the primordial stars.}


\titlerunning{Rotational effect on primordial minihalos}
\maketitle

\section{Introduction}
\label{sec:introduction}

Since the late 1960s, the theoretical study of the Universe at high 
redshift has persuaded a number of groups to work persistently on 
the dynamics of collapsing gas clouds 
\citep{sz67,pd68,silk77,cr86,htl96,tegmark97}, leading to the formation 
of the very first stars in the Universe, the so-called Population~III 
(or Pop~III) stars. Based on these leading$\hbox{-}$edge studies, 
the first sources of light are believed to have formed only a few 
hundred million years after the Big Bang \citep{bl01,bl04,cf05}. This 
era marks a crucial transition from the simple to the complex Universe 
\citep{loeb10,by11,g13}. Subsequent pioneering three-dimensional 
numerical simulations using adaptive mesh refinement \citep{abn00,on06}, 
smoothed-particle hydrodynamics \citep{yoh08} or more recently 
using the hydrodynamic moving mesh code {\em Arepo} 
\citep{gswgcskb11,gbcgskys12} have led to the development of a widely 
accepted standard model of primordial star formation. In this star 
formation  model, the first protostars form within dark matter (DM) 
halos with a virial temperature of $\sim 1000$ K and masses of 
$\sim 10^5\hbox{--}10^6$ $M_\odot$, which had collapsed at 
redshift $z \geq$ 20.
However, other simulations show that Pop~III stars may still form 
well beyond this, in waves that delay their formation \cite[e.g.,]
[]{ssf03,tfs07,ritter12}. These studies suggest that Pop~III star 
can continue to form down to the redshift $z = 2.5$ with a low peak 
rate of 10$^{-5}$ $M_{\odot}$ yr$^{-1}$ Mpc$^{-3}$, which occurred 
at $z = 6$. Observational evidence of Pop~III stars and the possibility 
of Pop~III waves reaching lower redshifts has received a significant 
boost with the discovery of a luminous Lyman-$\alpha$ (Ly$\alpha$) 
emitter with high EW He {\sc ii} and Ly$\alpha$ emission and no 
metal lines \citep{sobral15}.

The hydrogen atoms combine with the free electron (present from the 
epoch of recombination at $z \sim$ 1100) to produce small fractional 
abundances of $\rm H_2$, $x_{\rm H_2} \sim 10 ^{-3}$ \citep{suny98,yahs03}. 
As the collapse proceeds, the gas is cooled via $\rm H_2$ rotational 
and vibrational line emission (also termed as gas-phase reaction), 
with $\rm  H^-$ ion as an intermediate state (first discussed in the 
context of the local ISM by \citealt{McDowell1961})
\begin{equation}
\rm H + e^- \rightarrow H^- + \gamma,
\end{equation}
\begin{equation} 
\rm H^- + H \rightarrow H_2 + e^-,
\end{equation}
where the free electrons act as catalysts \citep{on98}. At this 
point, gas attains a temperature $\sim$ 200 K, and is in local 
thermodynamical equilibrium (LTE) with the kinetic temperature of 
the gas. However, the limited $\rm H_2$ abundance is not sufficient 
to cool the gas further, and the gas begins to heat up with increasing 
density. This results in the free-fall time being shorter 
than the cooling time. The transition from the cooling (low density) 
to heating (high density) with increasing density occurs near a 
critical density $n_{\rm cr} \approx 10^4$  cm$^{-3}$, and sets 
a characteristic Jeans length, allowing the gas to fragment with 
the Jeans mass of $M_{\rm J}$ (200\,K, $10^4\,$cm$^{-3}$) $\sim$ 
1000 $M_\odot$ \citep{abn02}. At higher density ($\sim 10^8$ cm$^{-3}$) 
hydrogen molecules are formed by the three-body reactions \citep{pss83} 
\begin{equation}
\rm H + H + H \rightarrow H_2 + H
,\end{equation}
\begin{equation}
\rm H + H + H_2 \rightarrow H_2 + H_2.
\end{equation}
There is however a significant uncertainty in the rate coefficients
\citep{gs09} for the above set of reactions that can cool the gas 
rapidly by converting almost all the atomic hydrogen into molecules, 
making the gas chemothermally unstable \citep{tao09,dutta15a}.

However, the amount of hydrogen molecules produced is a strong function 
of temperature. At high temperature, $\rm H_2$ molecules are destroyed 
by collisions with atomic $\rm H$ and $\rm H_2$ molecules \citep{yoha06},
\begin{equation}
\rm H_2 + H \rightarrow H + H + H,
\end{equation}
\begin{equation}
\rm H_2 + H_2 \rightarrow H + H + H_2.
\end{equation} 
The collision dissociation of $\rm H_2$ prevents the fractional 
abundance ($x_{\rm H_2}$) from becoming large. 
The study by \cite{tcggakb11} has discussed the equilibrium condition 
between the abundances of atomic and molecular hydrogen, known as the 
principle of microscopic reversibility. Therefore the rate at which 
$\rm H_2$ is produced via the three-body reaction must be 
compensated by the destruction rate to bring the system 
to a chemical and thermal equilibrium, the abundances of which are 
related through the following well-known Saha equation:
\begin{equation}
\frac{n_{\rm H_{2}}}{n_{\rm H}^{2}} = \frac{z_{\rm H_{2}}}{z_{\rm H}^{2}} 
\left(\frac{h^{2}}{\pi m_{\rm H} k T} \right)^{3/2} \exp \left(\frac{E_{\rm diss}}{kT}
\right),
\end{equation}
where $n_{\rm H_{2}}$ and $n_{\rm H}$ are the number densities of molecular 
and atomic hydrogen, respectively, $z_{\rm H_{2}}$ and $z_{\rm H}$ are the 
partition functions of molecular and atomic hydrogen, $E_{\rm diss}$ is the 
dissociation energy of the hydrogen molecule, $k$ is the Boltzmann 
constant, $h$ is Planck's constant, and $T$ is the temperature. 
In LTE, the adopted values of the rate coefficient for the $\rm H_2$
collisional dissociation ($k_{\rm diss}$) must be consistent with the 
three-body formation rate coefficient ($k_{\rm form}$) in the sense that 
each pair of rate coefficients satisfies the chemical equilibrium condition
$k_{\rm form}/k_{\rm diss} = K$, where $K$ is the equilibrium constant.

Once the gas density reaches a density $\sim 10^{10}$ cm$^{-3}$, 
the cloud becomes optically thick to the strongest of $\rm H_2$ 
lines. Using the Sobolev approximation (as described in \citealt{yoha06}), 
the $\rm H_2$ cooling rate in this regime can be expressed as
\begin{equation}
\Lambda_{\rm H_2, \rm thick} = \sum\limits_{u,l} h\nu_{ul} \beta_{\rm esc, ul} A_{\rm ul} n_{\rm u} \, ,
\end{equation}
where $n_{\rm u}$ is the number density of the hydrogen molecules in 
upper energy level $u$; $A_{\rm ul}$ is the spontaneous radiative 
transition rate, also known as Einstein coefficient, for a 
transition between $u$ and $l$; $h\nu_{ul}$ is the energy difference 
between $u$ and $l$; and $\beta_{\rm esc, ul}$ is the escape probability 
associated with this transition, i.e., the probability that the emitted 
photon can escape from the region of interest. In the high-density 
regime ($\sim 10^{14}$ cm$^{-3}$), the gas goes through a phase of 
cooling instability due to a strong increase in the cooling rate by 
$\rm H_2$ collisional induced emission (CIE) \cite[see, e.g.,][for 
detailed discussion]{ra04}. 
Above the central density $\sim 10^{16}$ cm$^{-3}$,
the gas becomes completely optically thick to the continuum radiation
\citep{yoh08}. At this point the remaining $\rm H_2$ dissociates
via reactions (5,6), and hence cools the gas to collapse further. 
Once all the $\rm H_2$ dissociates, the gas becomes fully adiabatic 
with core mass $\sim 0.01 M_\odot$ surrounded by a massive, dense 
envelope that accretes matter rapidly \citep{cgkb11a}. 

The vanguard numerical simulations \citep[e.g.,][]{abn02,yoha06,byhm09} 
propose the formation of massive (typically $\sim 20\hbox{--}50  
M_\odot$) primordial protostars. This  calculation result, however, 
contrasts with the present day star formation in which protostars with 
masses less than 1$M_\odot$ are formed \citep{kroupa02,chabrier03}. 
The recent improved and high resolution numerical simulations have 
inferred that the disk around the primordial core is unstable and 
fragments to form a small N system with low-mass stars, instead of 
a single protostar \cite[e.g.,][]{sgb10,cgsgkb11,hirano14,hcgks15}. 

In the literature, there are appreciable indications that the collapsing 
cloud from which the protostar forms could have strong rotational 
support \citep{larson69,larson84}. The cloud's rotation can affect the 
dynamical as well as the thermal evolution of gas and consequenty determine 
the ensuing properties of the Pop~III stars \cite[]{bodenheimer95,mhn97}. 
The consequences of the cloud's rotation on the chemical signature of 
the zero-metallicity primordial stars have been studied using stellar 
models \cite[][and references therein]{meynet09}. With the use of the sink 
particle technique (discussed in \S \ref{sec:gadget2}) in the smoothed
particle hydrodynamics (SPH) simulations, \citet{sbl11} have discussed
the rotation velocity of the first stars, angular momentum transfer and 
the internal structure of the new-born protostars \cite[see also][for 
the {\em Arepo} simulations]{sgkbl13}. The central protostar rotates 
with a significant fraction of the Keplerian velocity. There is however 
scatter in the radial velocity, temperature, and accretion rate. In a 
recent study, \cite{hirano14} performed radiation hydrodynamical 
simulations to follow the evolution of 100 primordial protostars. 
Although their simulations were in 2D, they nicely compared the angular 
momentum of the cloud with the Pop~III accretion rate. More recently, 
simulations and analytic models have shown the formation of the massive 
primordial stars in rapidly rotating disks in the presence of turbulence 
and UV backgrounds \cite[][and references therein]{latif13}. However, 
the extent to which the thermal and dynamical evolution of gas 
depends on the initial degree of rotation of the cloud has never been 
systematically tested. In addition, there is so far no clear 
understanding of how the cloud's rotation can regulate the concurrent 
heating and cooling process during the collapse. 

The dependence of the resultant fragmentation on the 
cloud's initial rotation, however, has been shown in detail in previous 
studies \cite[e.g.,][]{momi2008,stm08}. After performing a number of 
idealized numerical experiments, these parameterized studies have 
concluded that the formation of either binary or multiple systems
depend highly on the initial rotation of the cloud. 
Nevertheless, these studies could not point out the influence on 
the thermal evolution of the primordial gas during collapse because 
their calculations adopted the model of equation of state (EOSs). In 
this work, we candidly scrutinize  the role of initial rotation of the 
collapsing gas on the heating and cooling process that controls the
chemothermal evolution of gas inside minihalos. In addition, we
perform rigorous calculations in both the idealized as well as more
realistic cosmological minihalos to thoroughly analyze the evolution 
of gas particles and their physical properties. This unique approach 
thus enables us to investigate in detail the thermal, chemical, and 
dynamical evolution of the baryonic matter in a full 3D simulation 
of Pop~III collapse to understand better the physical process and 
the resulting fragmentation behavior that occurs once the first 
object is formed. Finally, we address important issues, such as the 
relation of the physical property and accretion phenomenon of the 
protostars on the initial rotation of the collapsing core. We also 
discuss the possibility of survival of Pop~III stars until today.

The paper is organized in the following manner. In \S \ref{sec:gadget2} 
we describe the numerical setup of the simulations and the initial 
conditions. In \S \ref{sec:cooling} we briefly discuss the relevant 
physical concepts of the problem with an emphasis on the heating and 
cooling process that determines the temperature evolution. The details 
of the velocity structure are outlined in \S \ref{sec:velocity}. We 
discuss the accretion phenomenon comprehensively, followed by the 
implication of this study for the fragmentation of primordial gas in 
\S \ref{sec:fragmentation}. The long-term evolution of protostars 
are contoured in \S \ref{sec:sinks}. We summarize the main points and 
draw our conclusions in \S \ref{sec:summary}.

\section{Simulations}
\label{sec:gadget2}
In order to follow the gravitational collapse, one needs to ensure 
that the gas evolution in the simulations should not depend on the choice 
of minihalos. We therefore use two completely different numerical 
setups: minihalos from the cosmological simulations of \cite{gswgcskb11} 
obtained from the hydrodynamic moving mesh code {\em Arepo} 
\citep{springel10} and the artificial minihalos with an initially 
uniform density distribution of gas particles. In the following, we 
describe the initial condition and setup of the simulations.

To investigate the cosmological minihalos, we use snapshots at the 
epoch when the central number density is just below $10^6$ cm$^{-3}$, 
the onset of the crucial three-body $\rm H_2$ formation reaction. The 
mesh-generating points of {\em Arepo} can be interpreted as the 
Lagrangian fluid particles, which is the basic characteristic of the 
Gadget-2 SPH code \citep{springel05}. At this time, the 
minihalos contain masses of 1030 $M_{\odot}$ and 1093 $M_{\odot}$ 
with maximum central temperature of 469 K and 436 K, respectively. The 
complete physical nature of these minihalos (e.g., number density 
($n$), initial rotation ($\beta_0$), maximum and minimum temperature, 
etc) is summarized in Table~1. We use the {\em Arepo} output of these 
minihalos \cite[MH-1 and MH-2 from][]{gswgcskb11} as the initial 
conditions for our Gadget-2 implementation. Because of the conversion 
from the moving mesh to the SPH formalism \cite[see, e.g., ][]{sgcgk11,
dnck15}, we denote the minihalos as CH1 and CH2 for the Gadget-2 
execution. The numerical resolution for CH1 and CH2 in SPH Gadget-2 
simulations (for 100 SPH particles) is $\sim 10^{-2}$ $M_{\odot}$.  

For the artificial setup, we use the initial conditions that permit 
us to carry out a set of methodical numerical experiments. The randomly
distributed gas particles have been settled using the periodic boundary 
conditions in a box. During this initial phase, we keep the gas 
temperature fixed and do not follow the primordial chemical network. 
Once the gas particles are settled, we are in a position to 
perform the simulations for the collapse of gas due to self-gravity.

The gas particles are initially uniformly distributed in a spherical 
cloud of size $R_0 \sim 2.7$ pc and a total mass of $M = 2982$ $M_\odot$. 
The number density is $n = 10^{3}$ cm$^{-3}$ and the temperature 
$T = 200$ K. These initial conditions are equivalent to the primordial
gas clumps that collapse \cite[see, e.g.,][]{abn02,yoha06}. All clouds 
are modeled with 5 million SPH particles, and the mass of a single SPH 
particle is $m_{\rm SPH} = 5.9 \times10^{-4}$ $M_\odot$. Therefore,
the numerical resolution (roughly 100 SPH particles) is $0.06$ $M_\odot$. 
The free-fall time for the uniform density distribution is $t_{\rm ff} 
= \sqrt{3/32 \pi G \rho} = 1.37$ Myr and the sound crossing time is 
$t_{\rm sc} \approx $ 5 Myr. As the free-fall timescale is shorter 
than the sound-crossing timescale, the clouds immediately start to 
collapse under their self-gravity. The clouds are then given different 
degrees of solid body rotation and are not subject to the internal 
turbulent motions. The strength of rotational support can be described 
by the $\beta_0$-parameter \citep{sdz03}, 
\begin{equation}
\beta_0 = \frac{E_{\rm rot}}{E_{\rm grav}} = \frac{R_0^3\Omega^2}{3GM} ,
\end{equation}

\noindent where $\Omega$ is the angular velocity, and $E_{\rm rot}$ and  
$E_{\rm grav}$ are the magnitudes of the rotational and the gravitational 
energies, respectively. We perform ten different numerical experiments with 
$\beta_0$ = 0.0, 0.005, 0.007, 0.01, 0.02, 0.04, 0.05, 0.07, 0.1, 0.2.

\begin{table}
\caption{Physical properties of the cosmological minihalos are summarized:}
\label{tab:3bh2rates}
\begin{center}
\renewcommand{\footnoterule}{}  
\small
  \begin{tabular}{l l l}
\hline
    \hline               
Halo                &               CH1              &              CH2          \\
properties          &                                &                           \\ 
\hline
 $n$ (cm$^{-3}$)      &  10$^{6}$ (max) 71 (min)        &  10$^{6}$  (max)  85 (min) \\
 $T$ (K)              &  469 (max)      59 (min)        &   436 (max)       54 (min) \\ 
 mass ($M_{\odot}$) &  1030                          &  1093                     \\ 
 $n$-SPH              &  690855                        & 628773                    \\ 
 resolution ($M_{\odot}$)&  1.3 $\times 10^{-2}$     &  1.4 $\times 10^{-2}$     \\
$\beta_0$           &  0.035 (max)   0.025 (min)     &   0.042 (max) 0.03 (min)  \\
\hline 
\end{tabular}
\end{center} 
{\em notes}: $n$ denotes the number density, $T$ the temperature, $n$-SPH 
the number of SPH particles and $\beta_0$ the rotation parameter, 
respectively. The numerical resolution is calculated for the 100 SPH particles.
\end{table}

As the collapse progresses in the central region of the cloud, 
it is extremely difficult to simulate the higher density regime 
because of shorter timescale. To overcome this problem, we use the
sink particle technique in which the high-density region is replaced 
by a single sink particle with appropriate boundary conditions 
\cite[see, e.g.,][]{bbp95,kmk04,jap05}. The sink particle can then 
be assumed to be (or at least approximated by) a growing protostar.
The density threshold for the sink particles to form is set as the 
number density of 5 $\times 10^{13}$ cm$^{-3}$, at which point the 
gas has a temperature of $\sim$ 1000 K. The sink particle can 
accrete gas particles within its accretion radius $r_{\rm acc}$, which 
we fix at 6 AU,  the Jeans radius at the density 
threshold for sink creation. The corresponding Jeans mass for 
both the cosmological as well as the artificial clumps is 0.06 
$M_{\odot}$, so we can resolve both the cosmological and artificial 
minihalos. The softening parameter of the sinks is 1.2 AU. In 
order to avoid spurious formation of new sink particles out of 
the gas, the sink particles is prevented further from forming 
within $2\, r_{\rm acc}$ of one another.

We follow \cite{cgkb11a} for the implementation of the external 
pressure term and a time-dependent primordial chemical network.  
To model a constant pressure boundary, we used a modified 
version of the Gadget 2 momentum equation, 

\begin{equation}
\frac{d v_{i}}{d t} = - \sum_{j} m_{j} \left[
f_{i}\frac{P_{i}}{\rho_{i}^{2}} \nabla_{i} W_{ij}(h_{i})
+ f_{j}\frac{P_{j}}{\rho_{j}^{2}}\nabla_{i} W_{ij}(h_{j}) \right] , 
\end{equation}

\noindent where the contribution from the external pressure 
($P_{\rm ext}$) is subtracted from both $P_{i}$ and $P_{j}$ (i.e., 
$P_{i}$ and $P_{j}$ are replaced by $P_{i} -  P_{\rm ext}$ and 
$P_{j} -  P_{\rm ext}$, respectively). All quantities have their 
usual meaning. The chemical network includes primordial hydrogen, 
helium, and deuterium to model the chemical and thermal evolution 
of the metal-free gas inside minihalos. The details of all the 
chemical reactions are given in \citet{ga08} and references therein. 
We adopt the intermediate three-body $\rm H_2$ rate coefficient 
$7.7 \times 10^{-31} T^{-0.464}$ cm$^6$ s$^{-1}$ proposed by \cite{g08}. 

%
%
%
%
%
%
%
%
%
%
%


\section{Heating and cooling rate}
\label{sec:cooling}

\begin{figure*}
\centerline{
\includegraphics[width=2.46in]{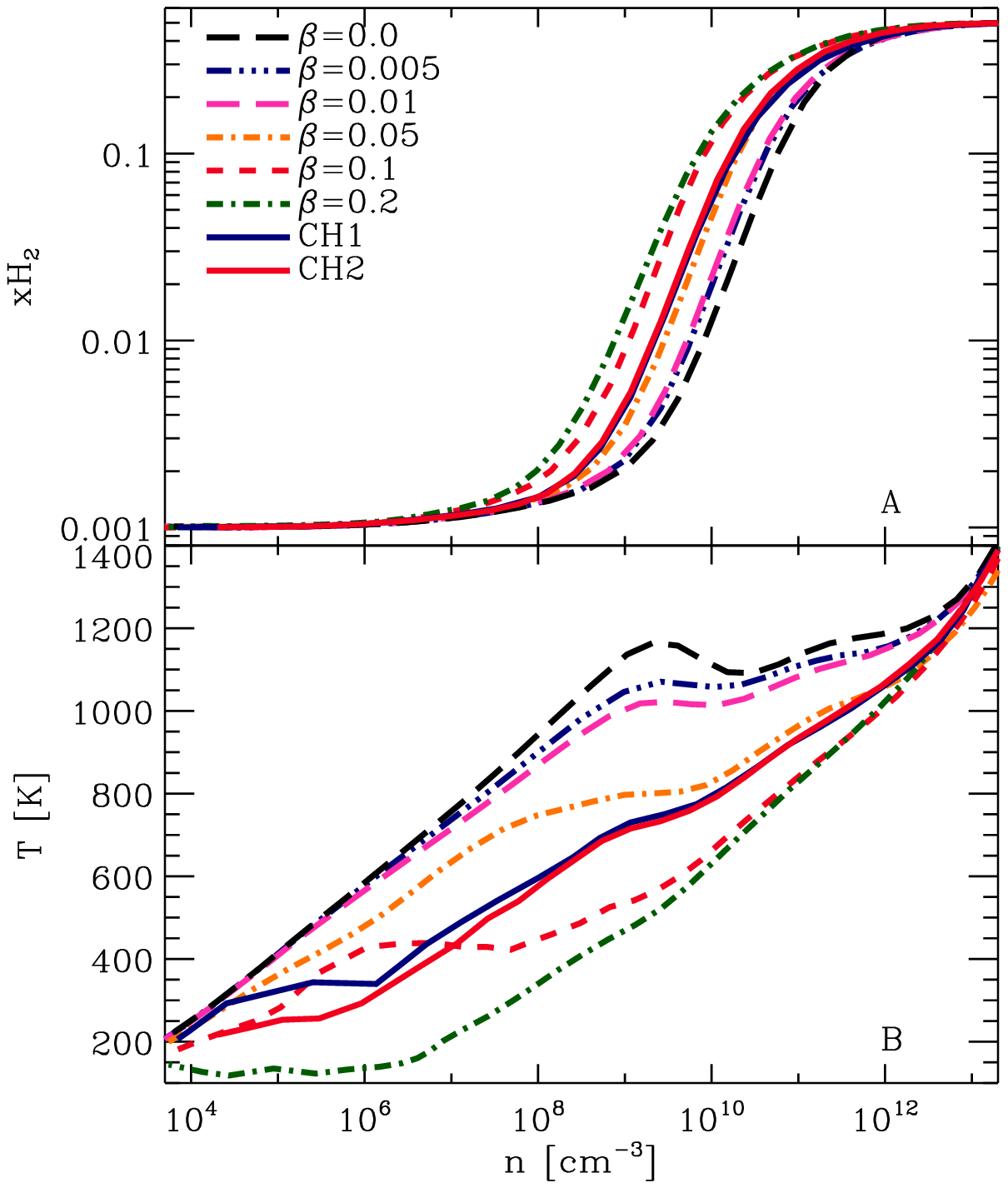}
\includegraphics[width=2.46in]{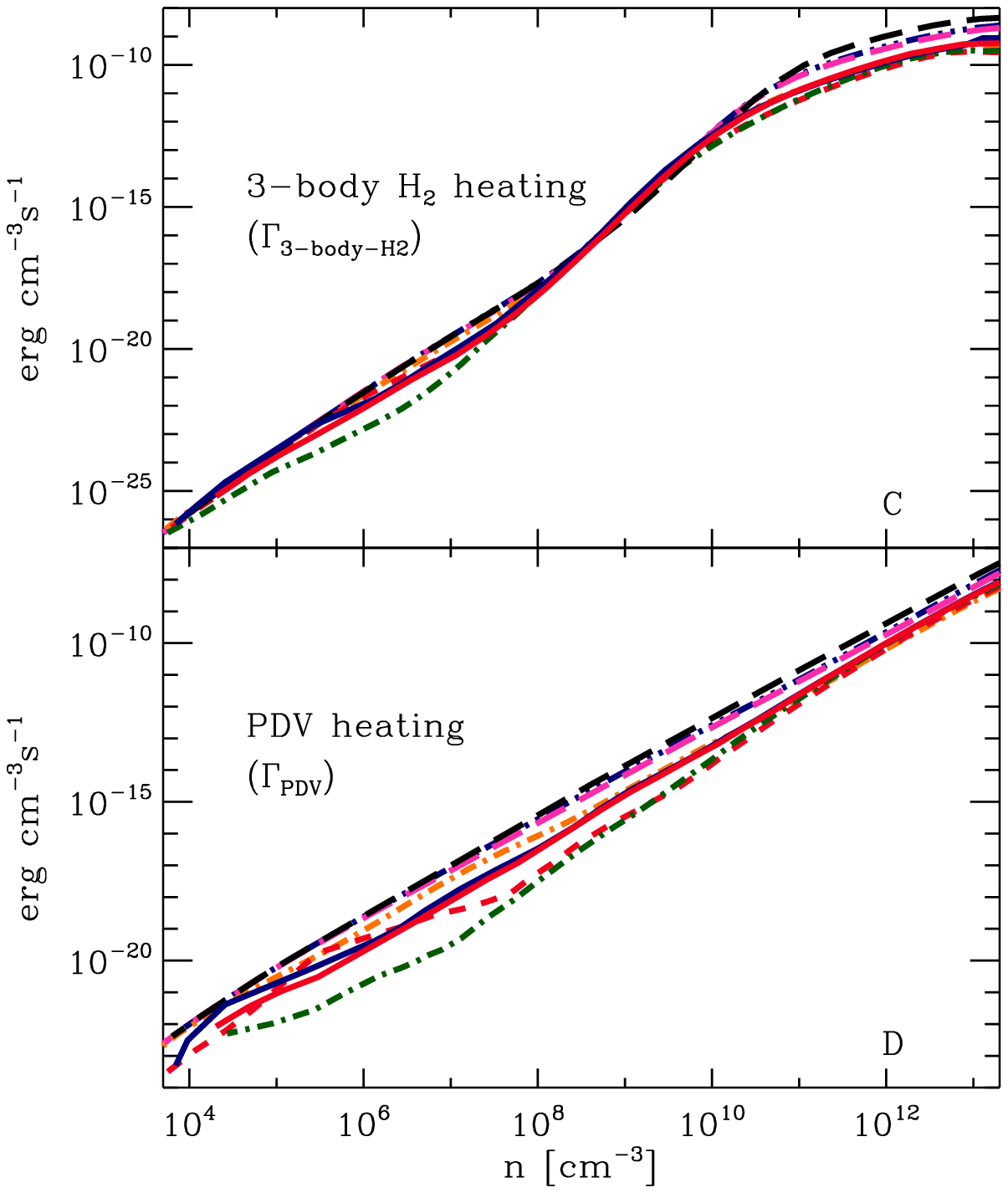}
\includegraphics[width=2.46in]{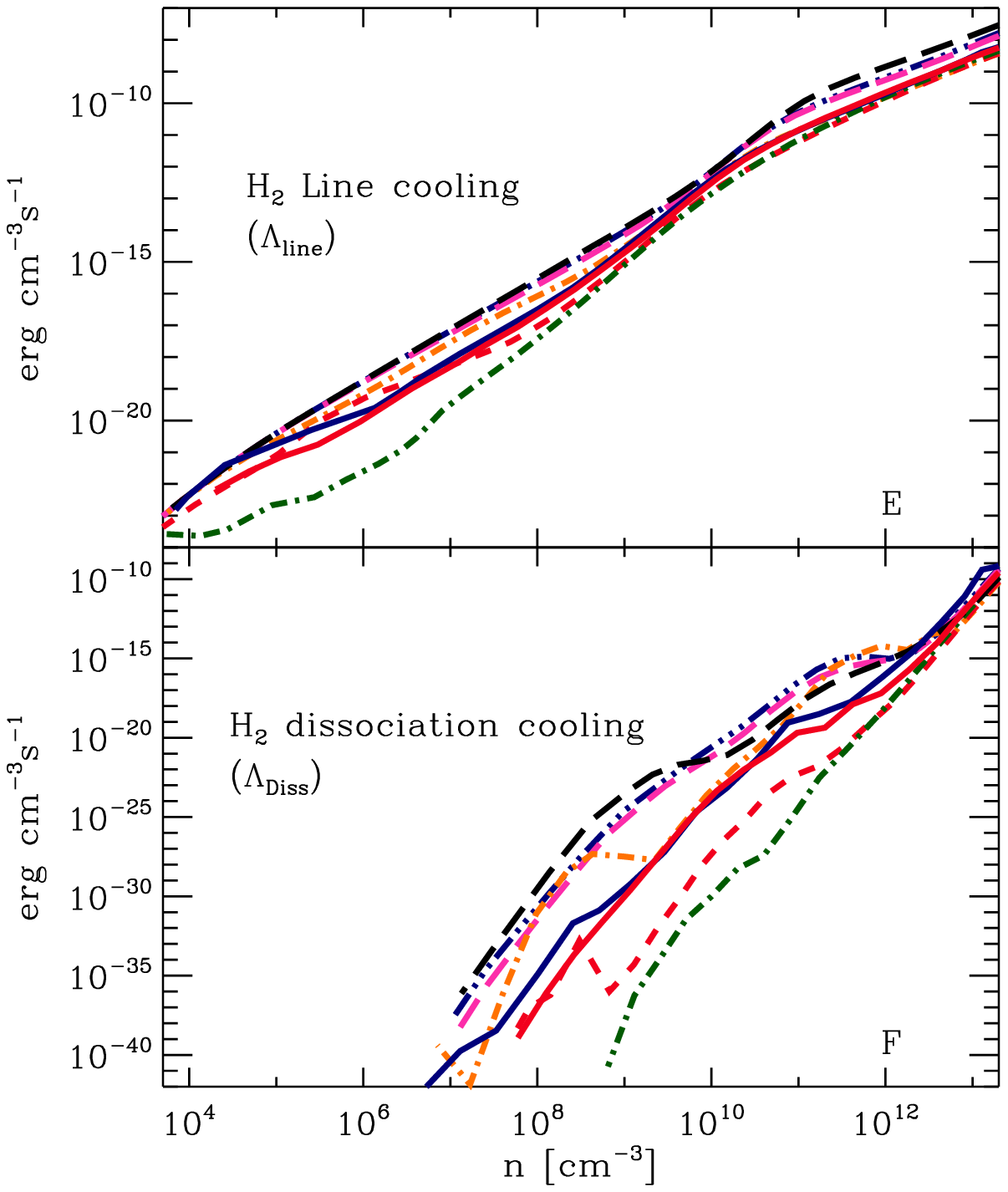}
           }
\caption{\label{heating} Radial logarithmic binned, mass-weighted averages 
of the $\rm H_2$ fraction (A), temperature (B), and various heating and 
cooling rates (C to F) are plotted as a function of density for different 
degrees of initial rotation $\beta_0$, just before the formation of the 
first sink.  
}
\end{figure*}


In this section, we investigate the cooling and heating mechanism 
associated with the emission, chemical reactions and gas contraction 
during the collapse of the cloud under its own gravity. The differences 
in the cooling and heating rates can force the gas to choose different 
paths for its thermodynamic evolution. 

If we assume that the gas density ($\rho$) evolves with the free-fall 
time ($t_{\rm ff}$) of the gas, i.e., $d\rho/dt = \rho/t_{\rm ff}$
\cite[see, e.g.,][]{om2000,gs09}, the thermal evolution can 
then be followed by solving the energy equation
\begin{equation}
\frac{d\epsilon}{dt} = \frac{p}{\rho}\frac{d\rho}{dt} - \Lambda + \Gamma ,
\end{equation}
where $\epsilon$ is the energy per unit volume in the gas, and $\Lambda$ 
and $\Gamma$ are the cooling and heating rates, respectively, in units of 
erg s$^{-1}$ cm$^{-3}$. Figure~\ref{heating} shows the 
physical conditions in the gas once the central region has collapsed 
to a density of $\sim 5 \times 10^{13}$ cm$^{-3}$, i.e., just before 
the first sink formation. The panels show mass-weighted averages of the 
properties of individual SPH particles within the radial logarithmic bins.

The three-body $\rm H_2$ formation heating rate provides chemical 
heating associated with the release of 4.48 eV each time a $\rm H_2$ 
molecule forms. Because the $\rm H_2$-fractions are almost similar 
(Fig.~\ref{heating}A), there is no consequential effect of the 
three-body $\rm H_2$ heating on the cloud's initial rotation 
(Fig.~\ref{heating}C). However, we find that there are substantial 
temperature differences between the clouds (Fig.~\ref{heating}B). 
For example, the temperature of the cloud with $\beta$ = 0.2 is almost 
40-60\% lower than that with $\beta_0$ = 0.005. This is because 
a higher degree of the rotational support slows down the contraction 
and reduces the amount of compressional heating (Figure~\ref{heating}D). 
Thus, the cloud with $\beta$ = 0.2 has a temperature of roughly 
$T \leq$ 200 K due to a slower collapse and, hence, less efficient 
compressional heating, whereas the temperature is nearly 1100 K in 
case of $\beta$ = 0.0 or $\beta$ = 0.005.

The rapid conversion of atomic to molecular hydrogen during the 
three-body reaction cools the gas less than the free-fall time and 
hence causes chemothermal instability \citep{dutta15a}. At high 
densities the heating rate is as high as the line-cooling rate 
(Fig.~\ref{heating}E), however, again with slight differences over 
the range of $\beta_0$ modeled. At equilibrium, the formation rate 
is balanced by the dissociation, and hence the dissociation cooling 
rate behaves in the same way with density as the heating rate. The 
dissociation cooling rate varies between clouds, with that for 
$\beta_0$ = 0.005  nearly 10 orders of magnitude higher than 
for the $\beta_0$ = 0.2 model (Fig.~\ref{heating}F) in the 
density range where the three-body reaction dominates. We therefore 
conclude that it is indeed the compressional heating ($pdV$)  
 that determines the thermal evolution of gas, which strongly depends 
on the initial degree of rotation ($\beta_0$). 

At this point, it is worth pointing out that the temperature 
evolution of the cosmological minihalos are similar to that of the 
idealized cases. As expected, both CH1 (with $\beta_0 \approx$ 0.035) 
and CH2 (with $\beta_0 \approx$ 0.042) have the temperature variation 
that falls in between the highest and lowest $\beta_0$ modeled throughout 
the density space. This also confirms that our varied parameter study 
with idealized clumps actually represents the cosmological initial
conditions of the minihalos.

\section{Velocity structure}
\label{sec:velocity}

\begin{figure*}
\centerline{
\includegraphics[width=2.46in]{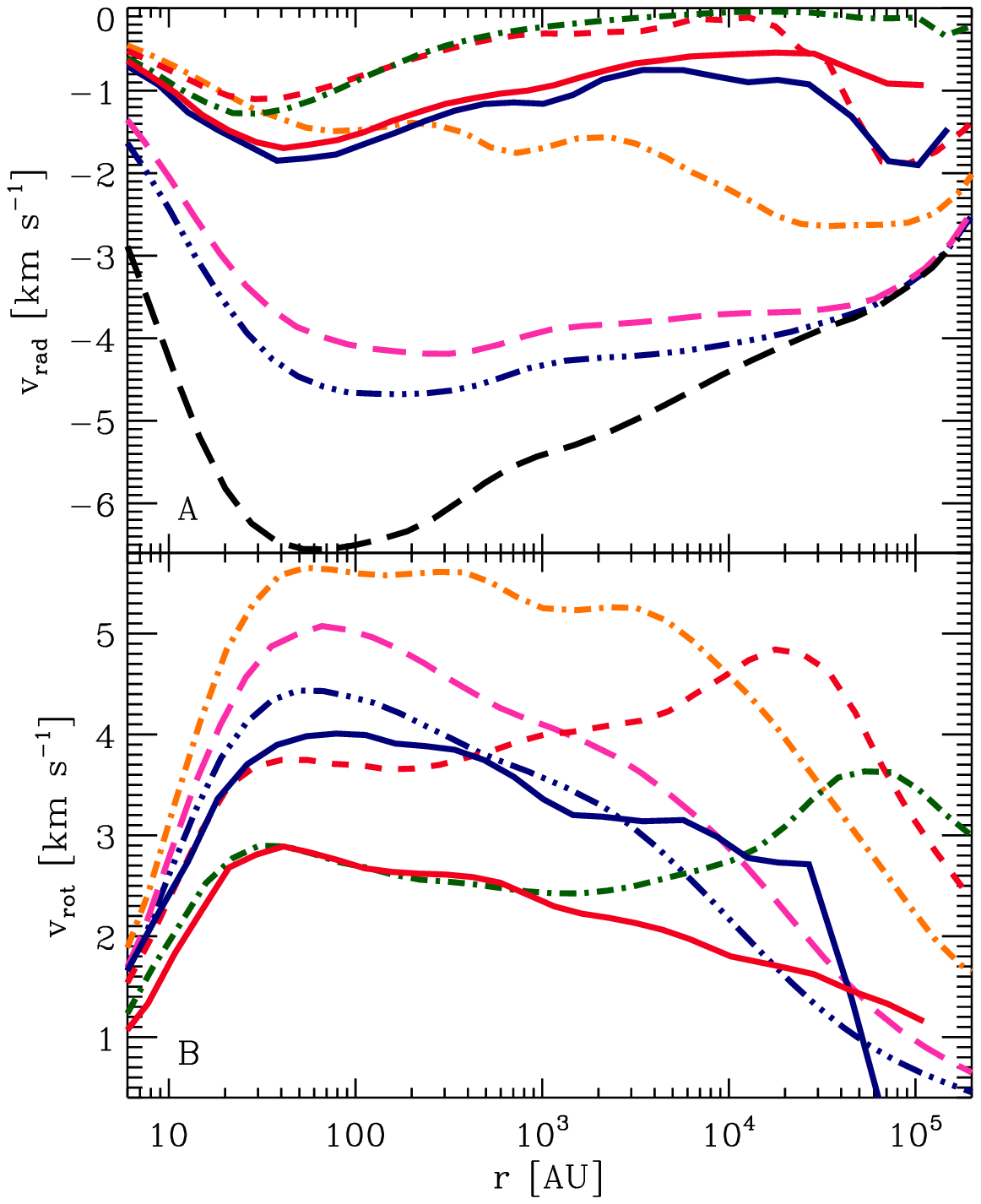}
\includegraphics[width=2.46in]{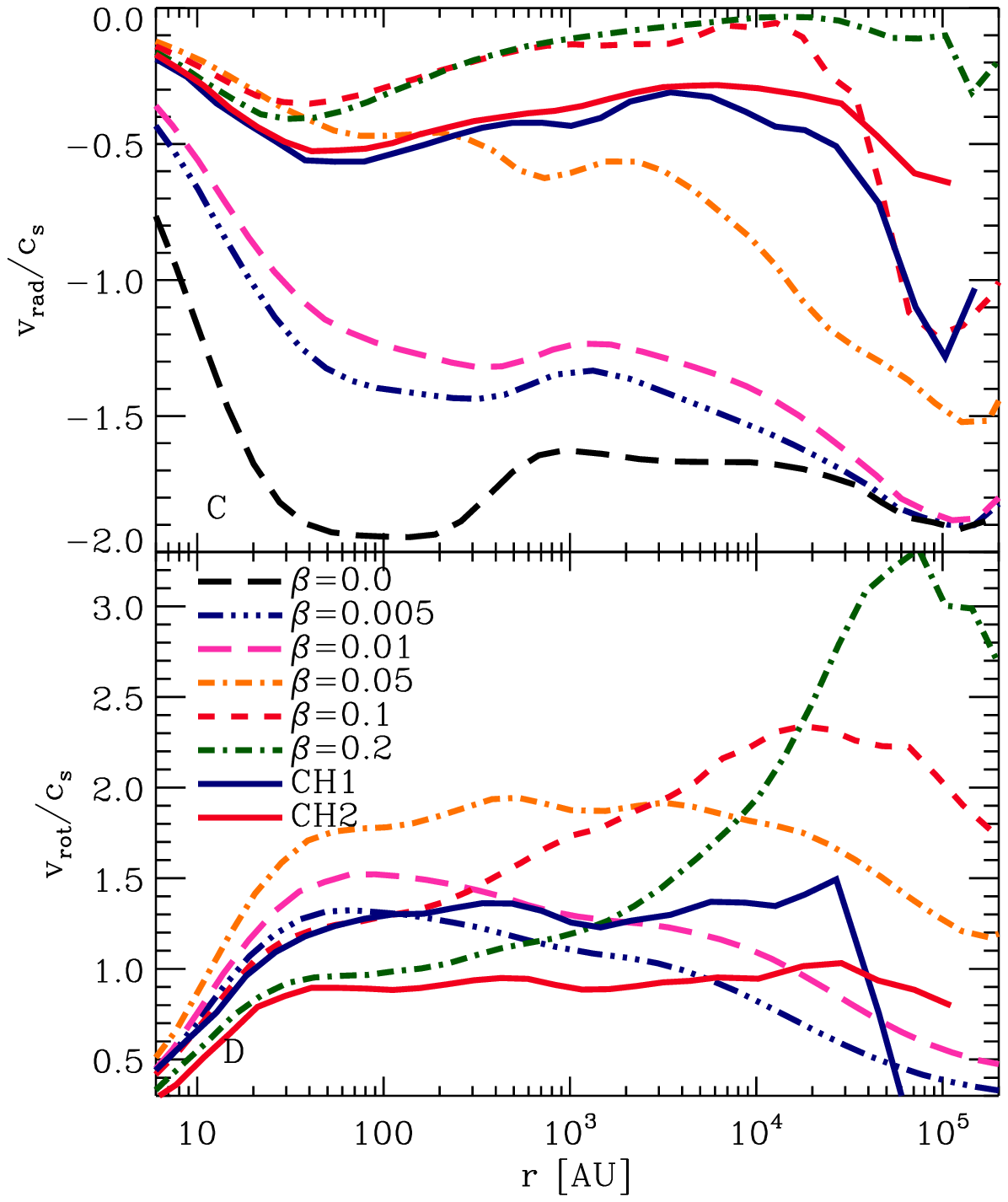}
\includegraphics[width=2.46in]{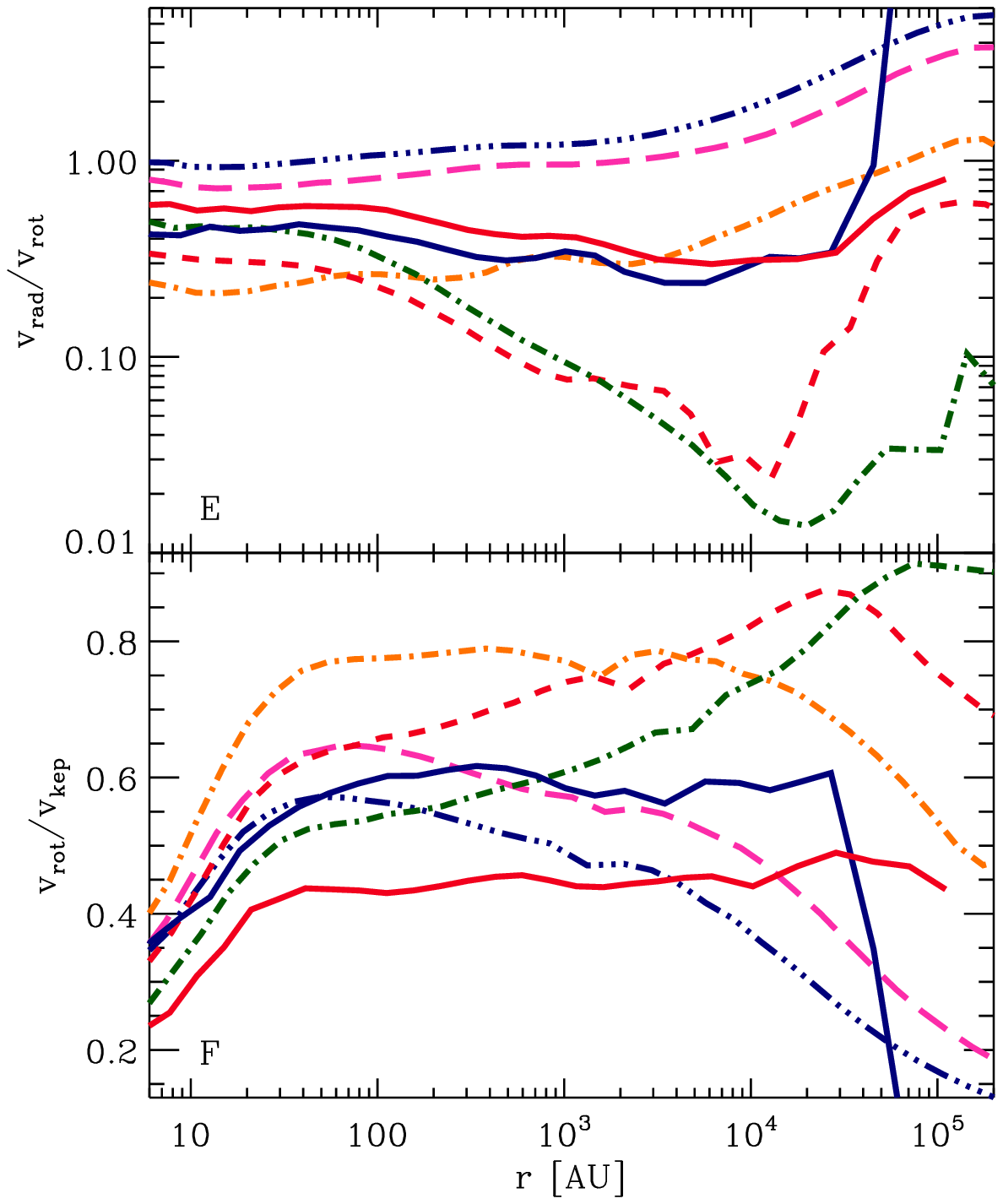}
           }
\caption{\label{vel} Radial logarithmic binned, mass-weighted 
averages of the radial velocity (A), rotational velocity (B), radial 
velocity over sound speed (C), rotational velocity over sound speed 
(D), rotational velocity over Keplerian speed (E), and radial velocity 
over rotational velocity are plotted as a function of radius, just 
before the first protostar forms. The initial strength of rotation 
introduces a scatter in the velocity.  
}
\end{figure*}


In this section, we study the dynamics of the gas particles that 
can arise as a result of the chemical and thermal evolution of the clouds. 
We therefore take a closer look at the velocity of the gas associated 
with the cloud collapse. The radial profiles of the gas show the 
mass-weighted averages within logarithmic bins and are taken just 
before the formation of the first sink.

We find that there are considerable differences in radial velocities 
between the clouds (Fig.~\ref{vel}A). The radial velocity of the 
cloud with $\beta_0$ = 0.005 is almost 30-40\% higher than the cloud 
with $\beta_0$ = 0.2. This is consistent with the fact that a lower 
rate of compressional heating for the gas of the collapsing core 
implies a lower radial velocity, which is nearly comparable with the 
sound speed (Fig.~\ref{vel}C). 

In order to quantify the degree of rotational support, we plot the 
rotational velocity (Fig.~\ref{vel}B) and the ratio of the rotational 
velocity ($v_{\rm rot}$) to the Keplerian velocity ($v_{\rm Kep}$), 
defined as $v_{\rm K} = \sqrt{GM_{\rm enc}(r)/r}$, where $M_{\rm enc}(r)$ 
is the mass enclosed within the radius $r$ (Fig.~\ref{vel}F). We 
find that the rotational speed for all $\beta_0$ modeled is well below 
that required for  rotational support by a factor of 5 to 7. This is 
consistent with previous cosmological simulations \cite[e.g.,][]{yoh08,
tao09}, which argued that the collapsing gas cloud has the ratio 
$v_{\rm rot}/v_{\rm Kep} \sim $ 0.5

For the $\beta_0 = 0.2$ case, the rotation 
speed varies between 0.3 and 0.9 times the Keplerian velocity, 
indicating that this cloud has gone through an efficient phase of 
angular momentum redistribution \citep{dutta15b}. However, for 
$\beta_0 >$ 0.05 the collapsing cloud is almost completely 
rotationally supported throughout and the gas at higher densities 
is relatively cold. All clouds are consequently sub-Keplerian, 
and the radial velocities for all clouds are comparable to the 
rotational velocities within 100 AU (Figure~\ref{vel}E). From
this discussion, we can infer that the cloud with higher rotation 
transfers the angular momentum more efficiently and hence becomes 
Keplerian in the outer regime of the collapsing core (for instance, 
$v_{\rm rot} \sim v_{\rm Kep}$ for $r \geq 10^4$ AU for $\beta_0 = 0.2$). 
Thus the outer regime, which is likely to form a Keplerian disk due to 
rotation, becomes unstable enough from accreting mass and, consequently, 
has higher chance to fragment (as we  see in the next section). We 
conclude that the cloud's initial rotation plays a pivotal 
role in the dynamical evolution of the gas by affecting the amount 
of rotational support even at later stages of the collapse.

\section{Mass accretion and fragmentation}
\label{sec:fragmentation}

During collapse, the angular momentum is transported to smaller 
scales, resulting in the formation of rotationally supported 
disk-like structures \citep{sbl11,gbcgskys12}. However, the 
rotation is not sufficient to hold the collapse of the disk, which 
then fragments into multiple objects \citep{cgsgkb11,dnck15}. We 
thus gauge the accretion and gas instability carefully, as 
predicted by the properties of the gas, to check whether there 
is any hint of fragmentation already present before the 
formation of the first sink. 

We follow \citet{abn02} to study the mass accretion rate, estimated 
as $\dot{M}(r) = 4 \, \pi \, r^2 \, \rho(r) \, v_{\rm rad}(r)$, as a 
function of radius for different degrees of initial 
rotation (Fig.~\ref{MBE}A). We also define the accretion time,
$t_{\rm acc} = M_{\rm enc}(r)/4 \, \pi \, \rho \, v_{\rm rad} \, r^2$.
For all simulations, $\dot{M}$  has a maximum at $\sim$ 
20-40 AU. Given that $\rm H_2$ line cooling becomes optically 
thick at the corresponding densities, $\dot{M}$ for all simulations 
converges in this range, as the gas looses its ability to cool 
efficiently. We would, however, like to point out that the scale, 
where $dM/dt$ becomes maximum, may slightly change with the choice 
of density threshold for the sink formation. For example, if we 
choose $n \sim 10^{15}$ cm$^{-3}$ for a sink particle to form, the 
collapse of the inner structure proceeds and hence the mass accretion 
rate reaches a maximum at $\sim$ 15-30 AU, depending on the initial 
degree of rotation. However, this is not a substantial issue 
compared to the overall thermal nature of the collapsing cloud. For 
slowly rotating clouds, the accretion rate is substantially higher 
($\sim 0.1$ yr$^{-1}$) and the accretion time is on the order of 
free-fall time (Fig.~\ref{MBE}C). 

Since the collapsing core becomes Jeans unstable by accreting more 
and more mass, we check the strength of the gravitational instability 
by measuring the number of the Bonnor-Ebert masses ($M_{\rm {BE}}$) 
contained in the central dense volume \citep{ebert1955,bonnor1956}. 
We compute the Bonnor-Ebert masses as 
$M_{\rm{BE}} = 1.18 (c_s^4/G^{3/2})P_{\rm{ext}}^{-1/2} \approx
20 M_\odot T^{3/2} n^{-1/2} \mu^{-2} \gamma^2$ \citep{abn02}, where 
$c_s$ is the sound speed, $P_{\rm{ext}}$ is the external pressure that 
we assume to be equal to the local gas pressure, $\mu$ is the mean 
mass per particle, and $\gamma = 5/3$ is the adiabatic index, respectively. 
Within the central 10$^4$ AU regime, the enclosed gas mass for all 
values of $\beta_0$ contains a roughly equal number of Bonnor-Ebert 
masses, although with a factor of two between the highest and lowest 
values of $\beta_0$ (Fig.~\ref{MBE}B). However, in the outer
regime (i.e., $r \geq$ 10$^4$ AU), the clouds with $\beta_0=0.1$ and
$\beta_0=0.2$ contain roughly 7 and 10 Bonnor-Ebert masses respectively.
This is because the higher rotating clouds become close to Keplerian
in the outer regime (as seen in the previous section) and tend to 
form a disk-like structure. As the higher rotation obstructs the
infalling gas particles, the disk gradually accumulates enough 
Jeans masses through accretion and becomes gravitationally unstable to 
fragmentation.

Alongside the Bonnor-Ebert analysis, we also compare all relevant 
timescales associated with the cloud collapse. Figure~\ref{MBE}C
represents the accretion timescale ($t_{\rm acc}$) over free-fall 
timescale ($t_{\rm ff}$). For a higher rotating cloud, the accretion 
time is much longer than the free-fall time. This is consistent with 
the above discussion and an alternate way to explain the mass 
accretion features of clouds with different $\beta_0$-models shown in 
Fig.~\ref{MBE}A. In Figure~\ref{MBE}D, we consider the fragmentation 
timescale, defined as $t_{\rm frag} \equiv M_{\rm BE}/\dot{M}$ 
\cite[e.g.,][]{dgck13}, and compare it with the free-fall timescale. 
We know that fragmentation generally
occurs when the dynamical timescale of the central collapse becomes 
larger than the collapse timescale of individual density fluctuations.
Here the free-fall timescale represents the dynamical timescale.
In the outer regime ($r \geq 2 \times 10^4$ AU), which is likely to
form a disk, the fragmentation time becomes shorter than the free-fall
time, especially for higher rotating clouds. We conclude here that 
the clouds with larger $\beta_0$ collapse slowly from high rotation
and tend to form a disk-like structure that becomes Keplerian and
gravitationally unstable by accreting mass, heralding fragmentation. 



%
%
%
%
%
%
%
%
%
%
%
%
%
\section{Protostellar system}
\label{sec:sinks}

\begin{figure}
\centerline{
\includegraphics[width=3.8in]{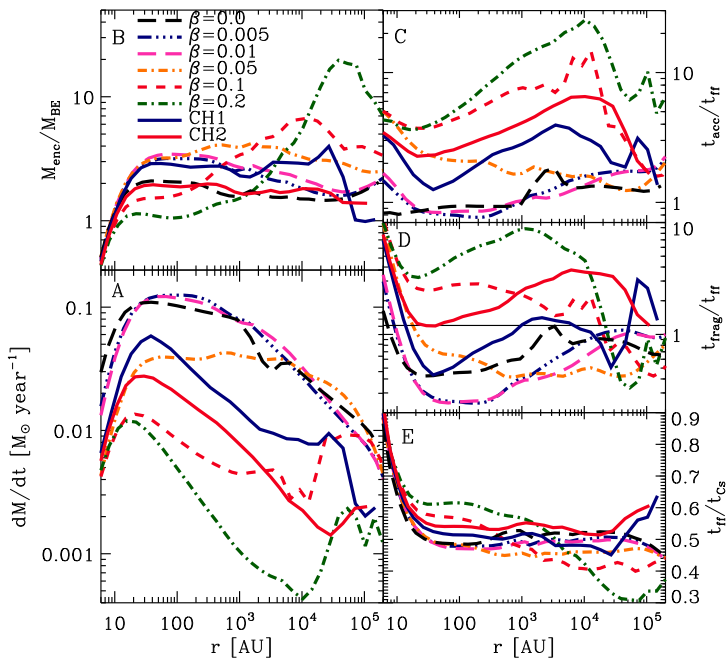}
}
\caption{\label{MBE} Radial logarithmic binned, mass-weighted 
averages of the mass accretion rate (A), the number of Bonnor-Ebert 
masses (B), the accretion time over free-fall time (C), the 
fragmentation time over free-fall time (D), and the free-fall time 
over sound crossing time (E) are plotted as a function of radius, 
just before the formation of the first sink.}
\end{figure}


\begin{figure*}
\centerline{
\includegraphics[width=3.55in]{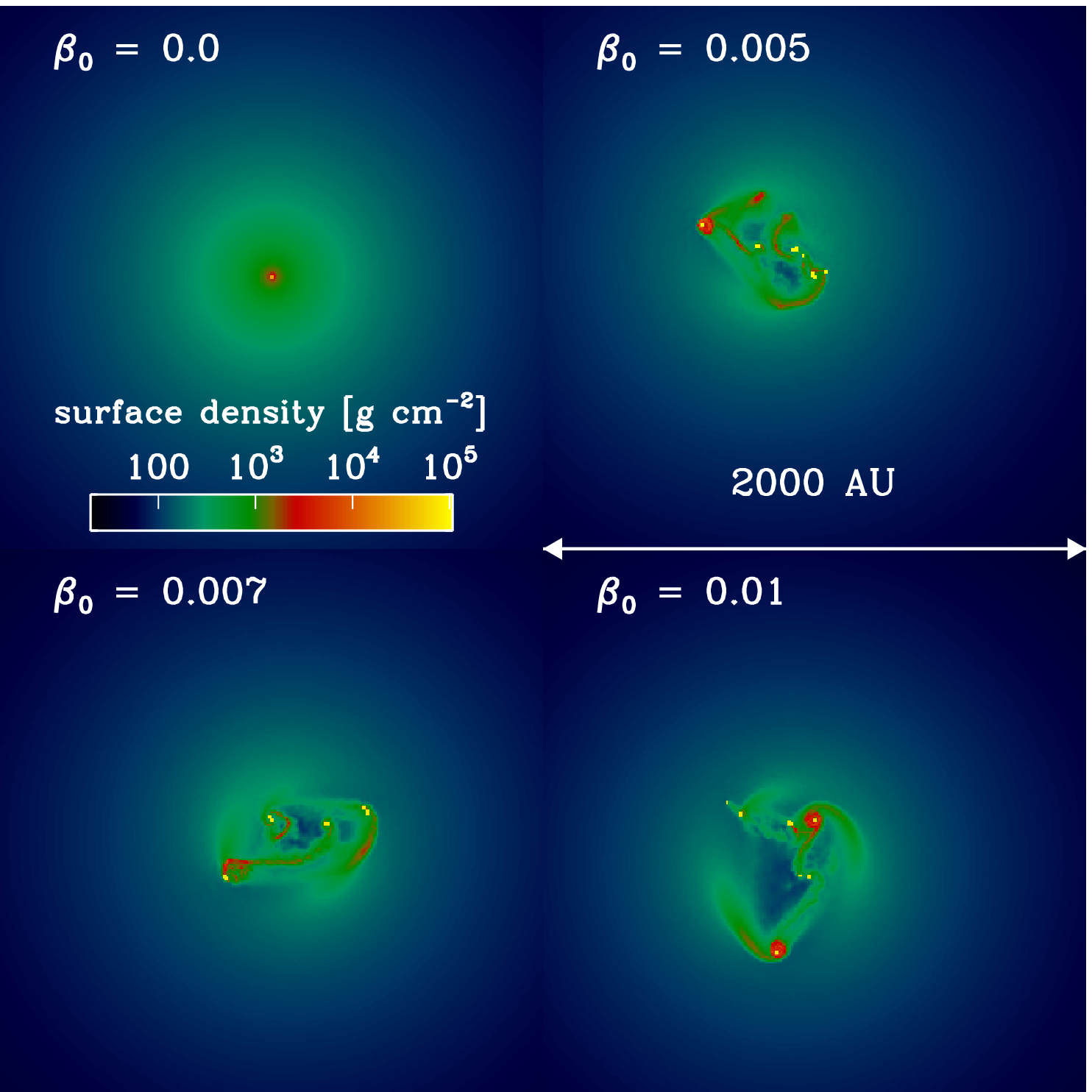}
\includegraphics[width=3.55in]{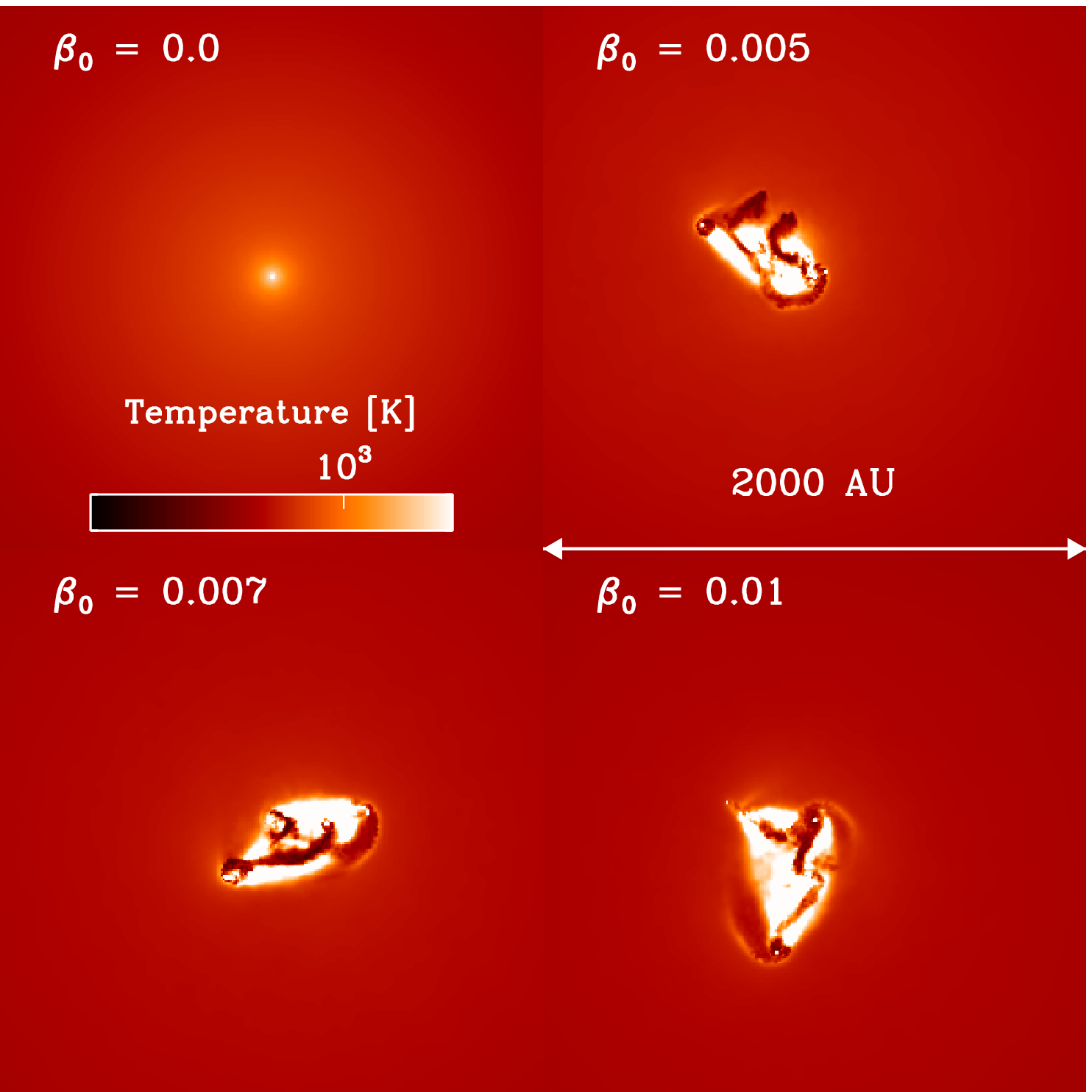}
}
\centerline{
\includegraphics[width=3.55in]{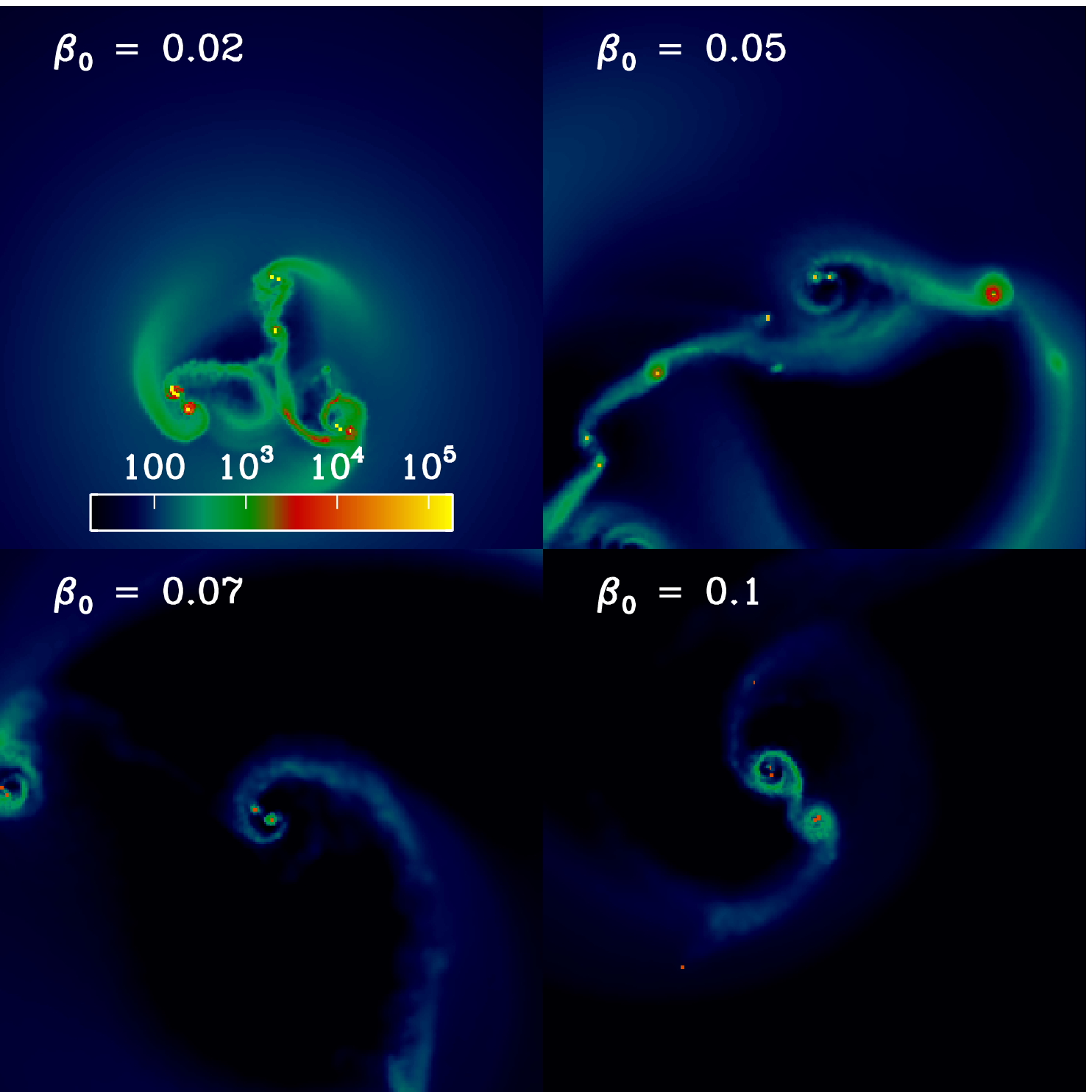}
\includegraphics[width=3.55in]{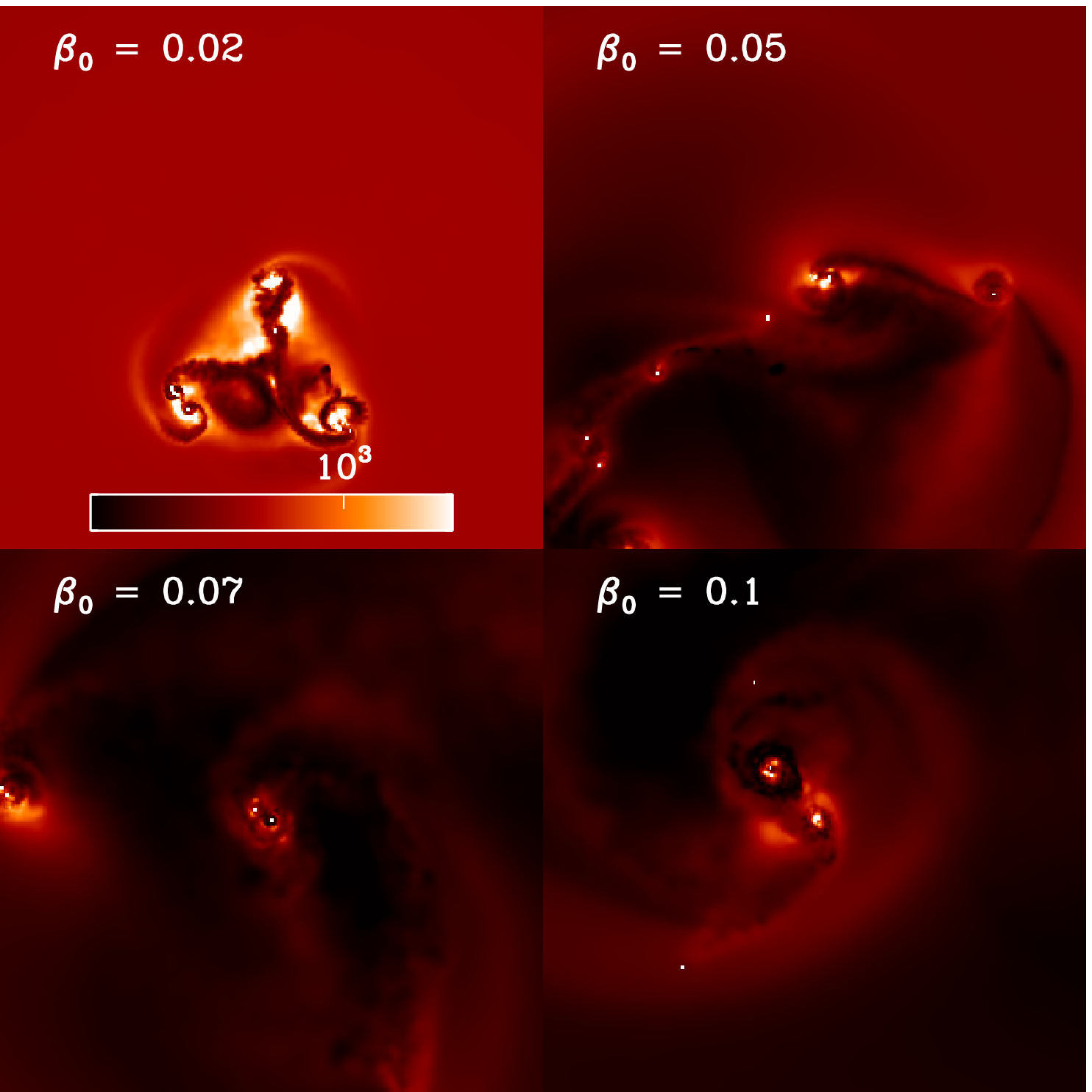}
}
\centerline{
\includegraphics[width=3.55in]{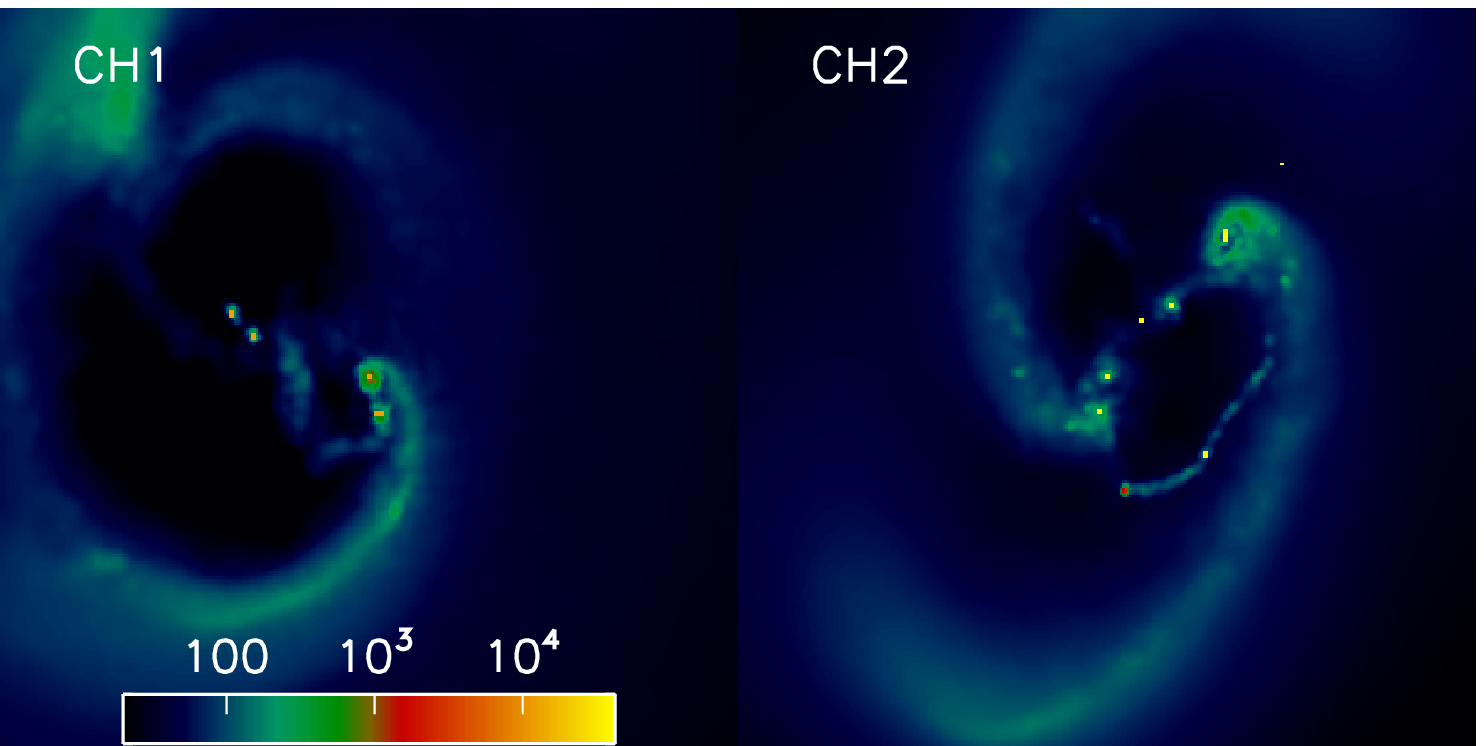}
\includegraphics[width=3.55in]{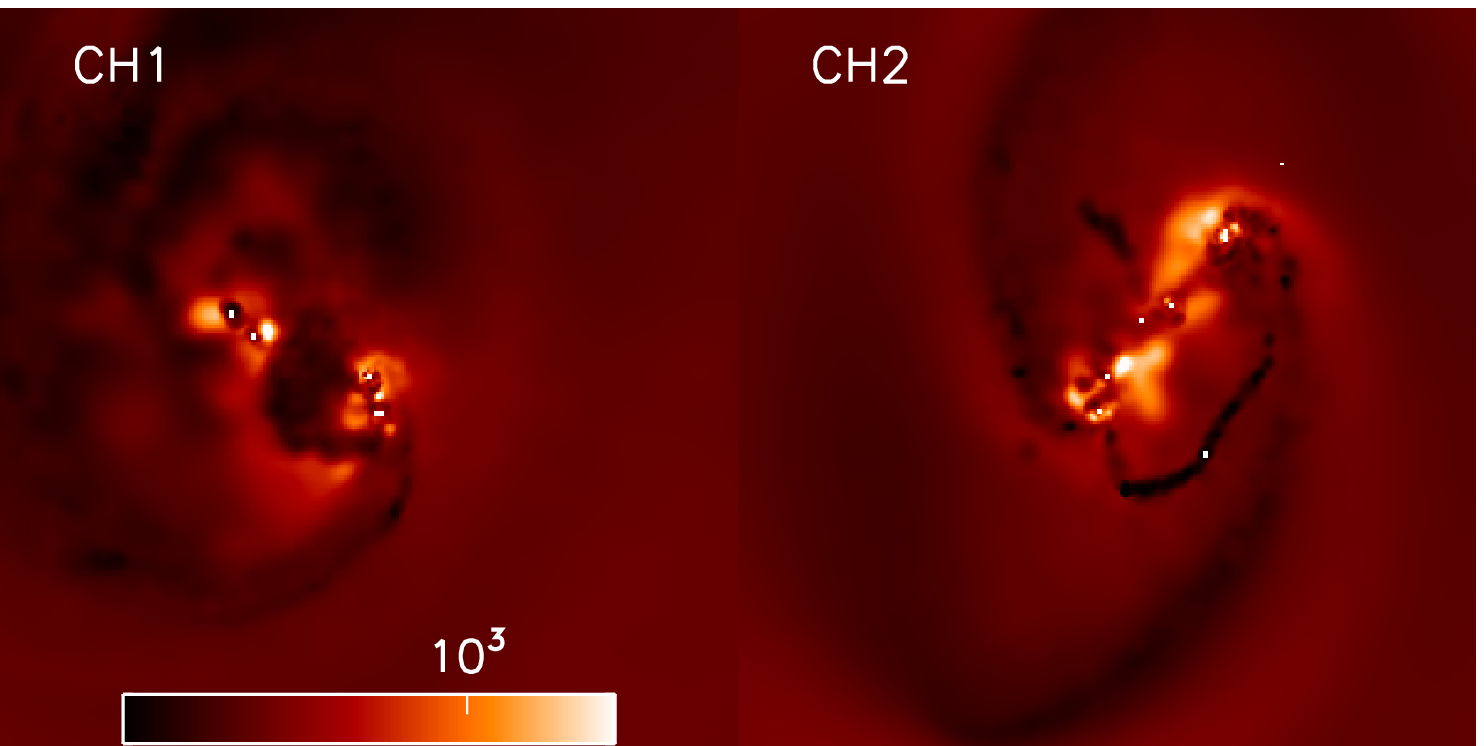}
}
\caption{\label{disk_image} The column density and column-weighted 
temperature distribution in a region of 2000 AU centered around the 
first protostar for different strengths of the initial rotation of 
the cloud are shown when a total of $\sim 30\,M_{\odot}$ have been 
converted into, or accreted onto, sink particles. 
}
\end{figure*}

\begin{figure*}
\centerline{
\includegraphics[width=2.46in]{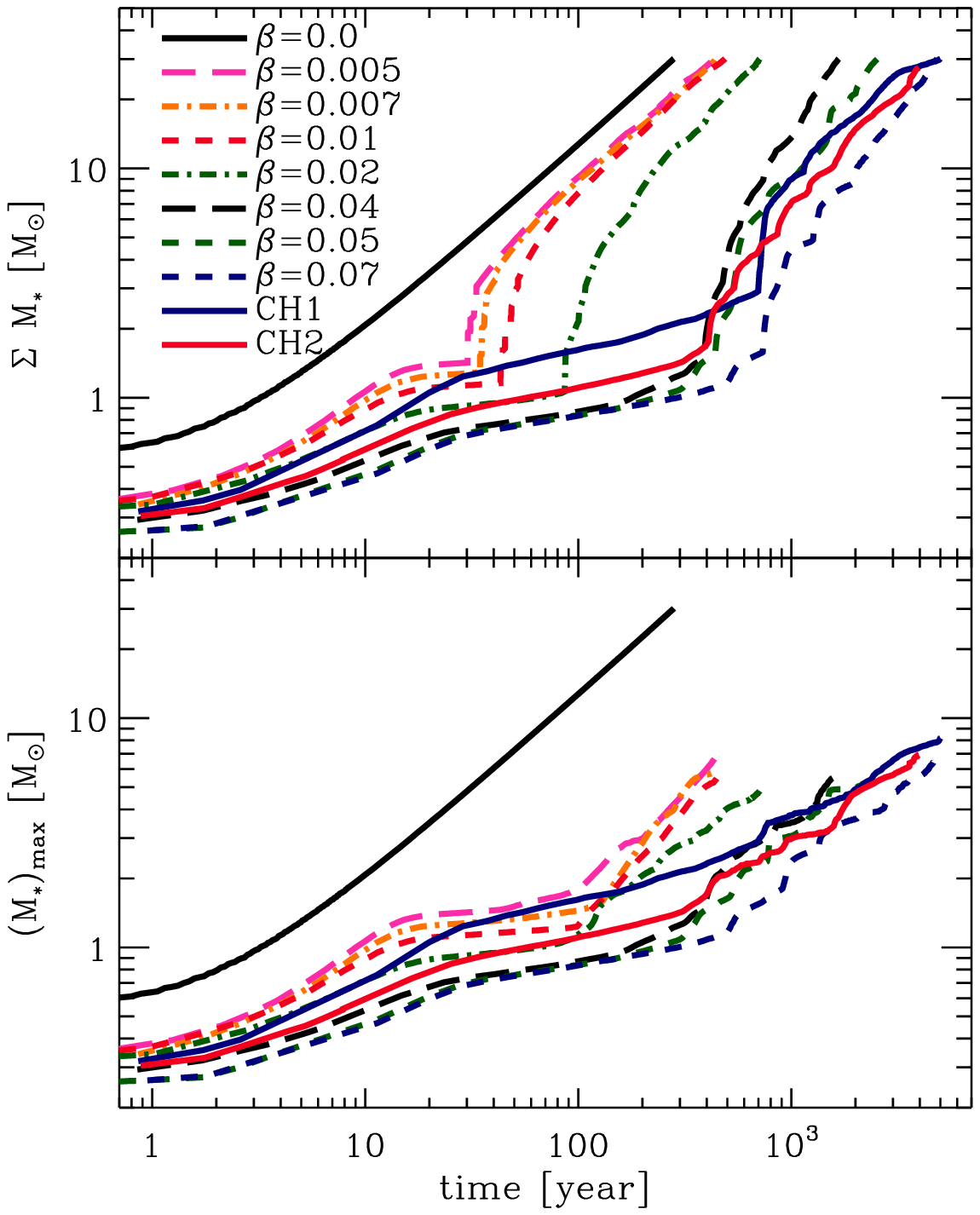}
\includegraphics[width=2.46in]{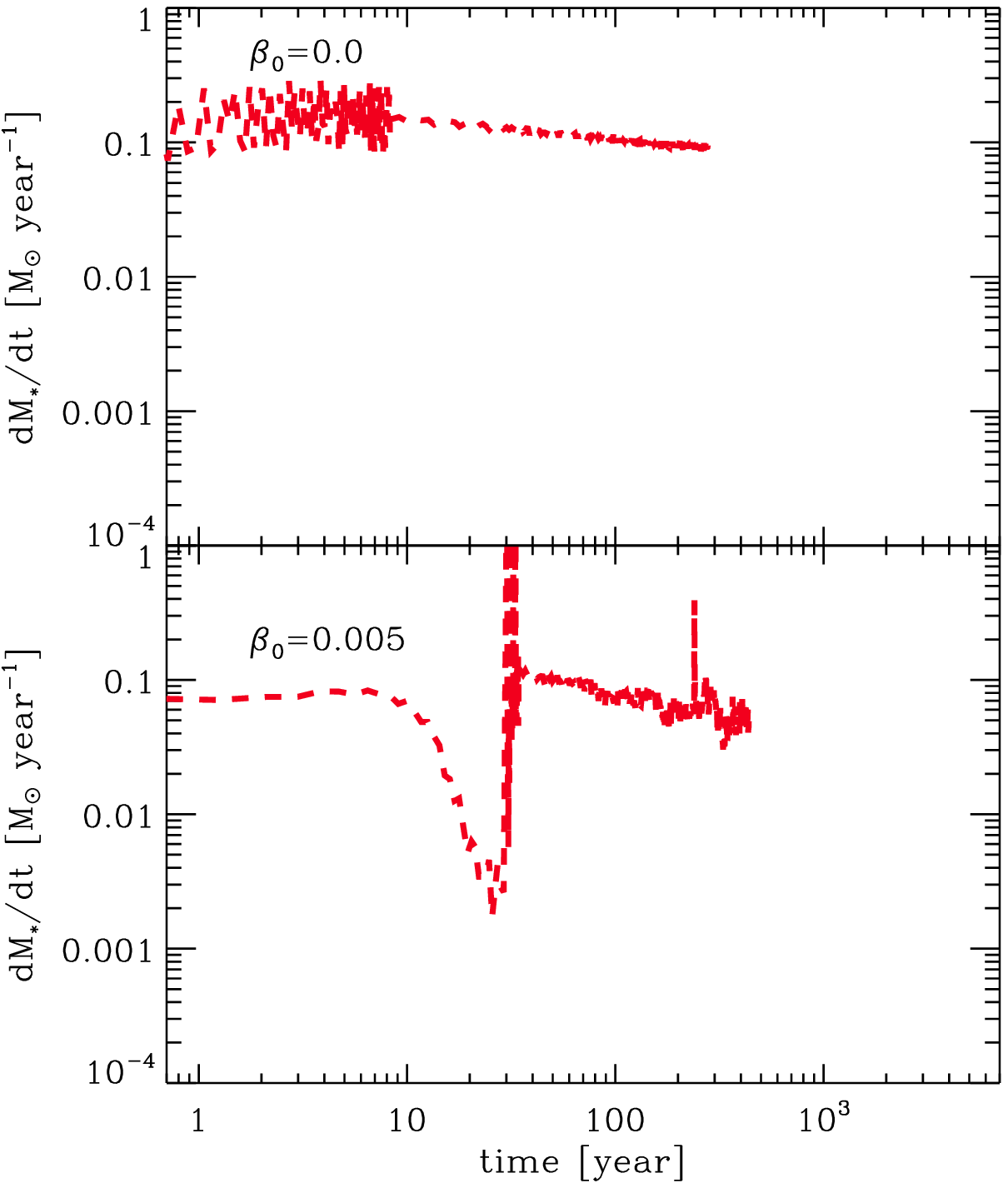}
\includegraphics[width=2.46in]{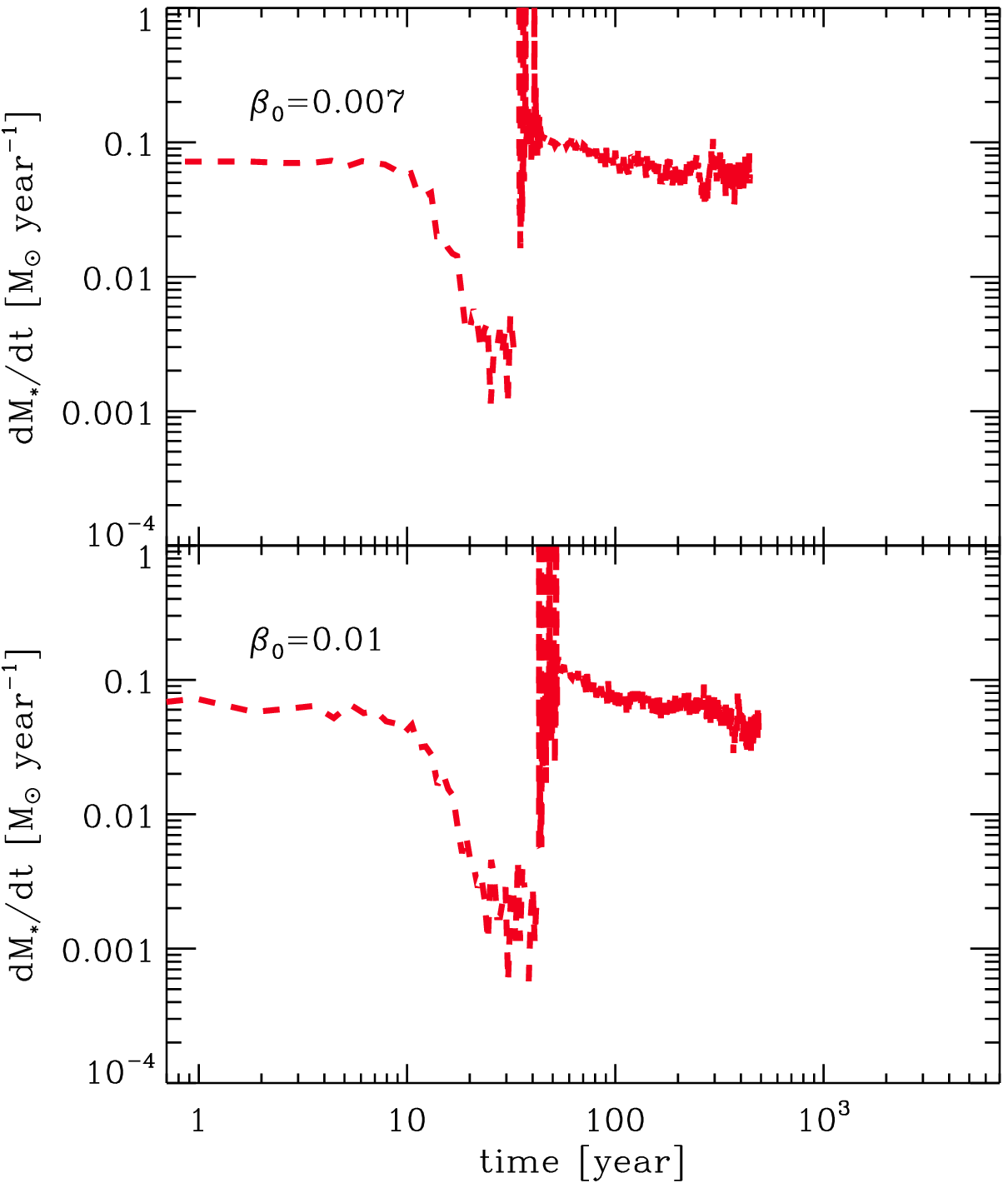}
}
\centerline{
\includegraphics[width=2.46in]{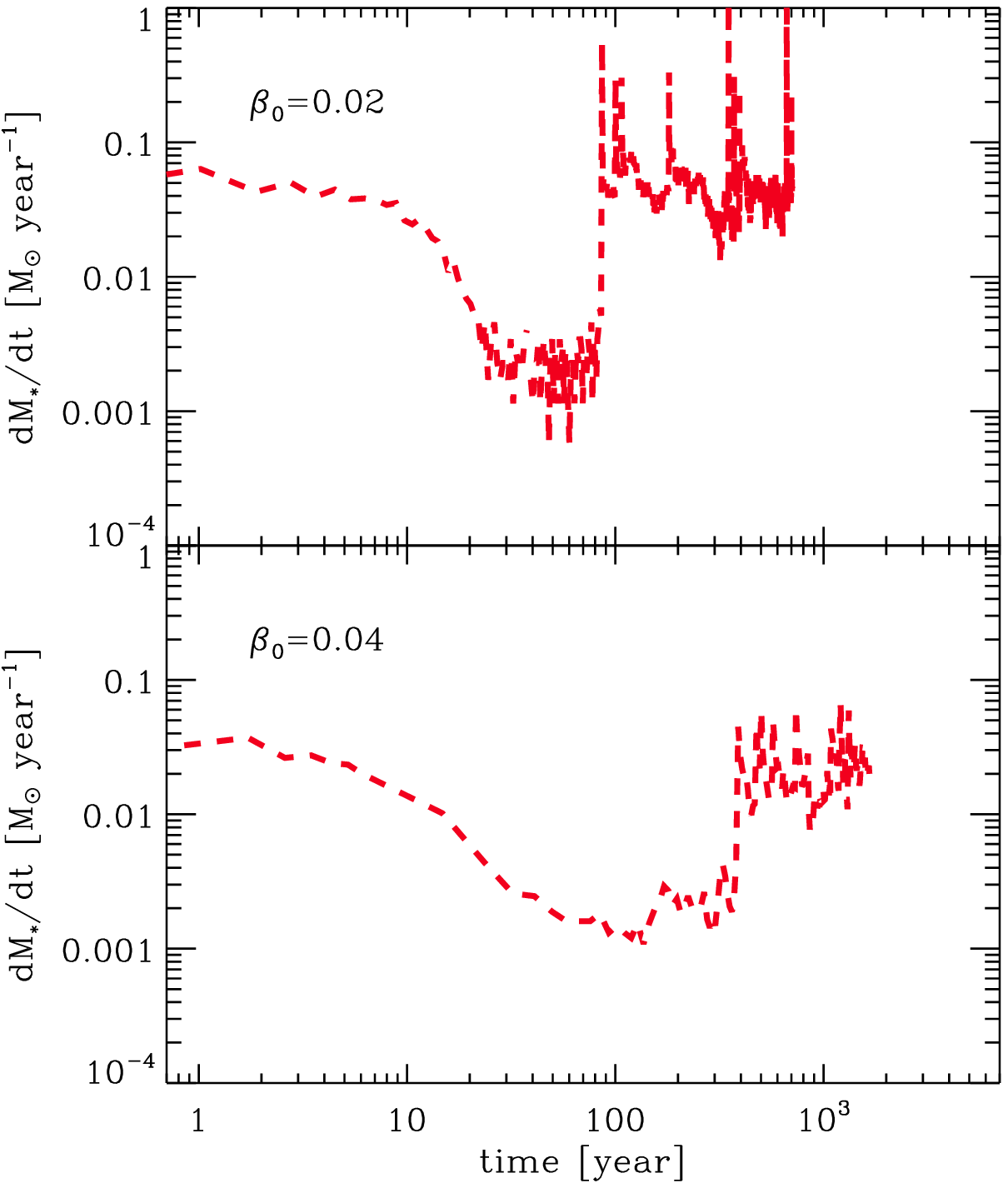}
\includegraphics[width=2.46in]{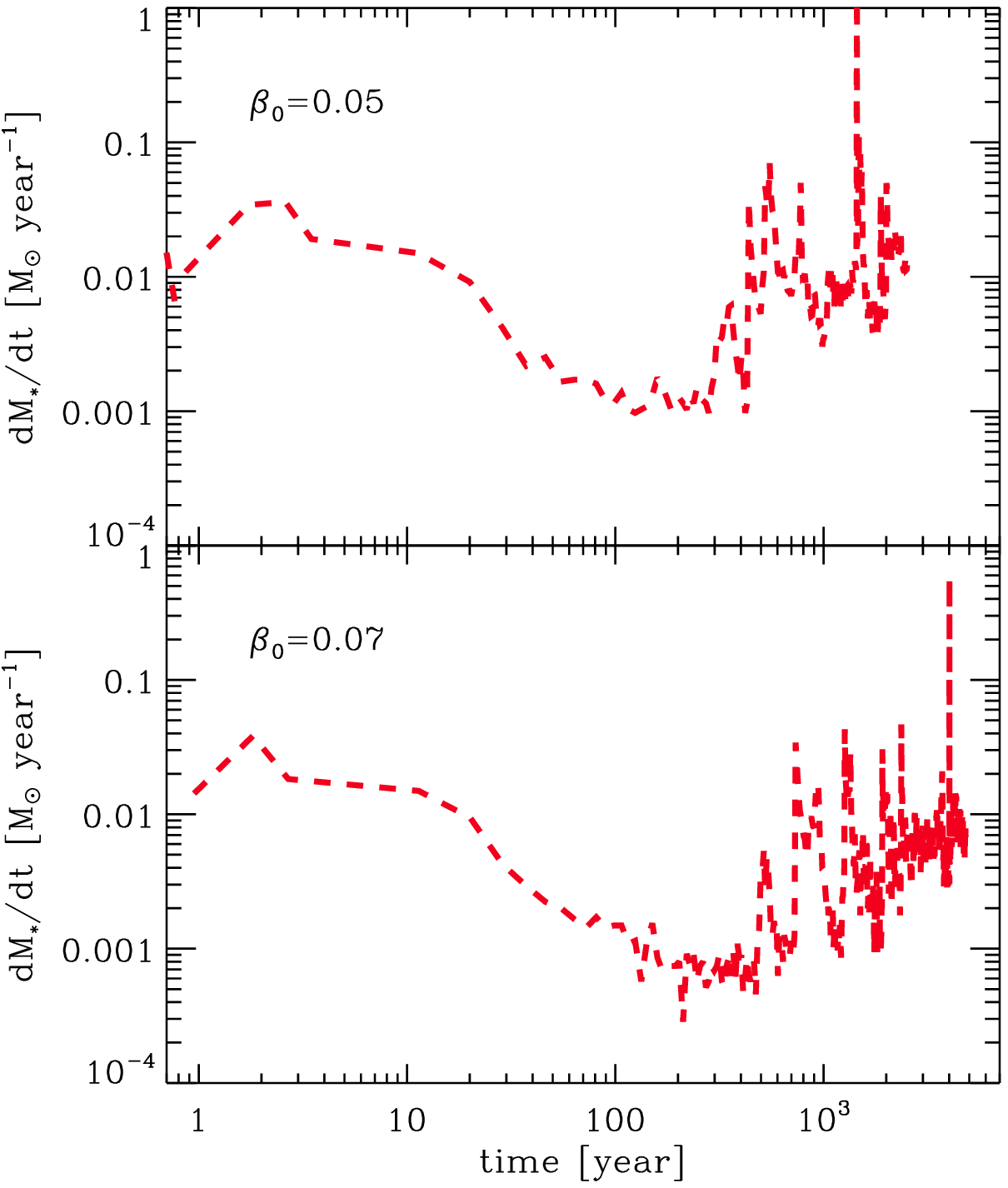}
\includegraphics[width=2.46in]{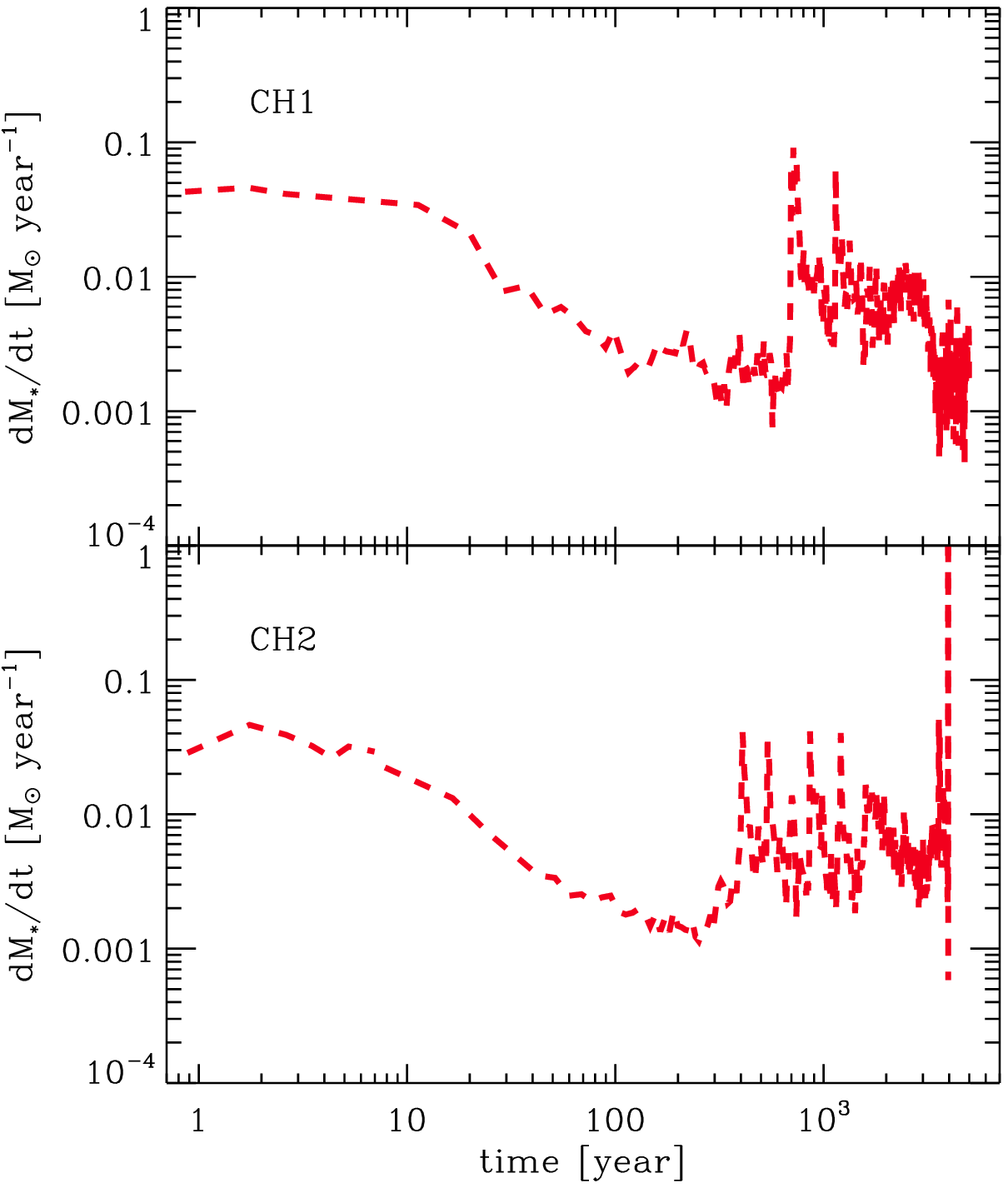}
}
\caption{\label{sinktime} Time evolution of the protostellar system: 
total mass of all the sinks particles and the most massive sink 
particle are plotted as a function of time (top left). Total mass 
accretion rate by all sink particles are shown as a function of time 
(for others rotation parameter $\beta_0$; same as Fig.~\ref{MBE}A). 
The cloud with zero rotation attains the 30 $M_{\odot}$ within a few 
hundred years after the formation of the first sink particle. The 
rotationally supported cloud takes $\sim$ 100 $\hbox{--}$ 1000 years, 
depending on the initial strength of rotation ($\beta_0$). The mass 
accretion rate decreases with time until further sink particles form. 
}
\end{figure*}


The study by \citet{momi2008} shows that any fragmentation only
takes place  after the disk formation, and that clouds with higher 
$\beta_0$ tend to fragment at lower densities. We examine how the 
circumstellar accretion disk that formed in the idealized, as well as
more realistic cosmological minihalos, becomes locally unstable and 
evolves for different degrees of initial rotation.


Figure~\ref{disk_image} shows the column density and column-weighted 
temperature distribution in the inner 2000 AU at the end of the 
simulations. These images clearly reflect that all simulations fragment 
to form a small N-body system, comprised of secondary protostars, 
within the time in which $\sim$ 30 $M_{\odot}$ of material are 
accreted onto the sink particles. It is not surprising that clouds 
with a higher level of rotational support exhibit the disk-like 
structure on several scales within the central region, whereas 
slowly rotating clouds are more likely to be centrally condensed. 

The top left panel of Fig.~\ref{sinktime} shows the time evolution 
of the total mass accreted onto all the sinks ($\Sigma M_*$) and the 
maximum sink mass ($M_{*\rm max}$) in the period over which the sink 
formation occurs. We find that the cloud with zero rotation attains 
the 30 $M_{\odot}$ within a few hundred years after the formation of 
the first sink particle, whereas rotationally supported clouds take 
$\sim$ 100 $\hbox{--}$ 1000 years, depending on the initial 
strength of rotation. Figure~\ref{sinktime} also shows the time 
evolution of the total mass accretion rate ($dM_*/dt$) onto all the 
sinks (same as Figure~\ref{MBE}A) for different values of $\beta_0$. 
As expected, the mass accretion rate is larger for the slowly rotating 
clouds. In each case, $dM_*/dt$ decreases with time until further 
sink particles form, and then the total accretion rate increases 
again, however, this time with large temporal variations. The mass 
accretion rate for the idealized cloud with $\beta_0=0.04$ 
is similar to that of the cosmological minihalo CH2, which  has a 
rotation parameter $\sim \beta_0=0.042$. This again confirms that 
our idealized clumps with varying $\beta_0$ can actually be the
representatives of cosmological initial conditions to investigate 
the thermodynamical evolution of gas and the resulting fragmentation 
behavior of the sinks.  

\begin{figure}
\centerline{
\includegraphics[width=1.6in,angle=270]{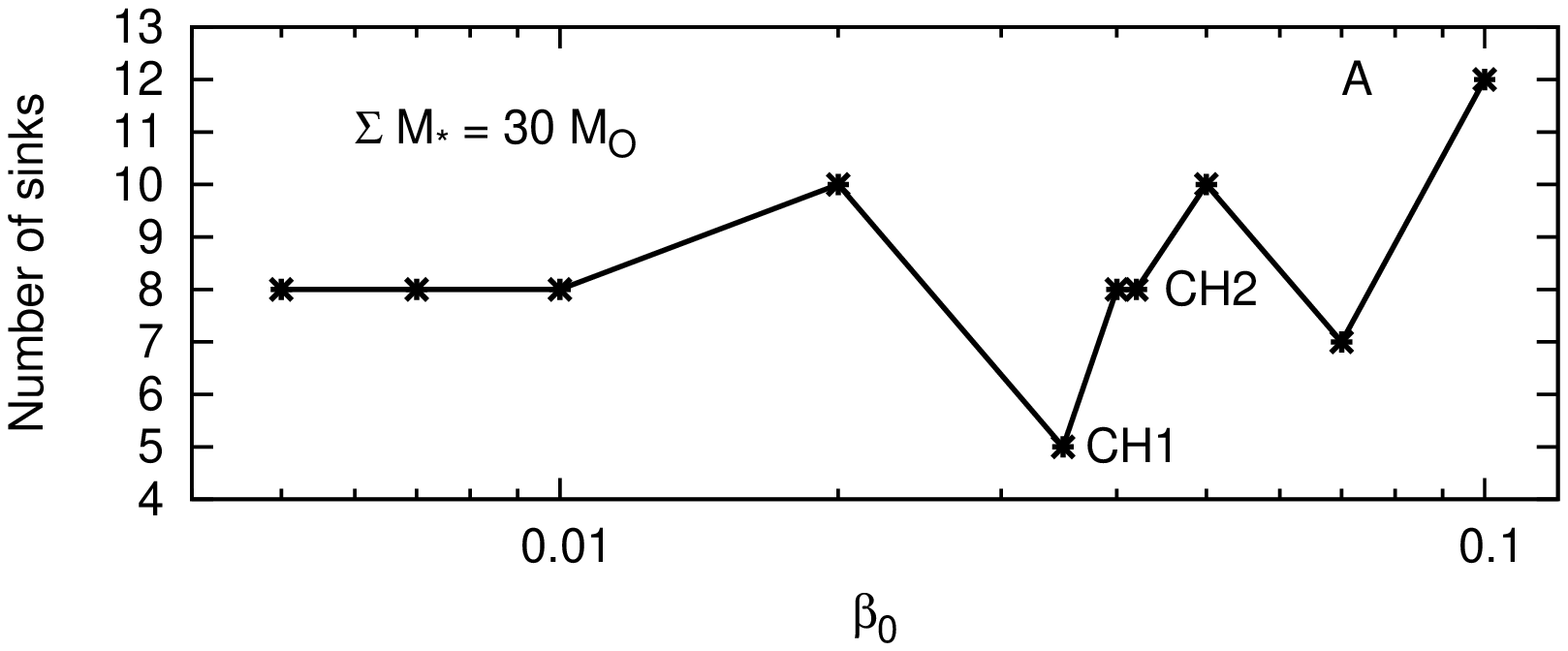}
}
\centerline{
\includegraphics[width=1.66in]{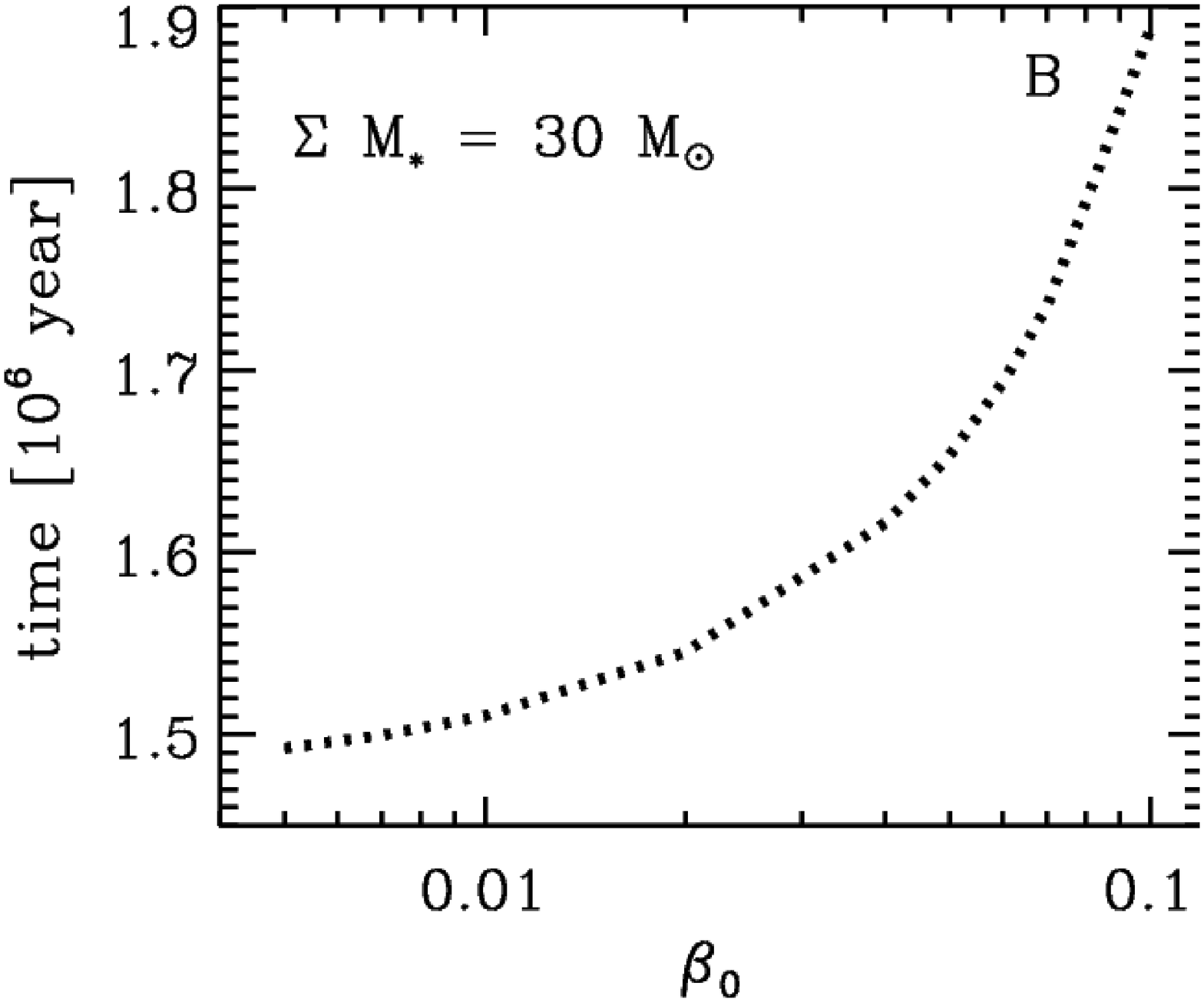}~~~
\includegraphics[width=1.66in]{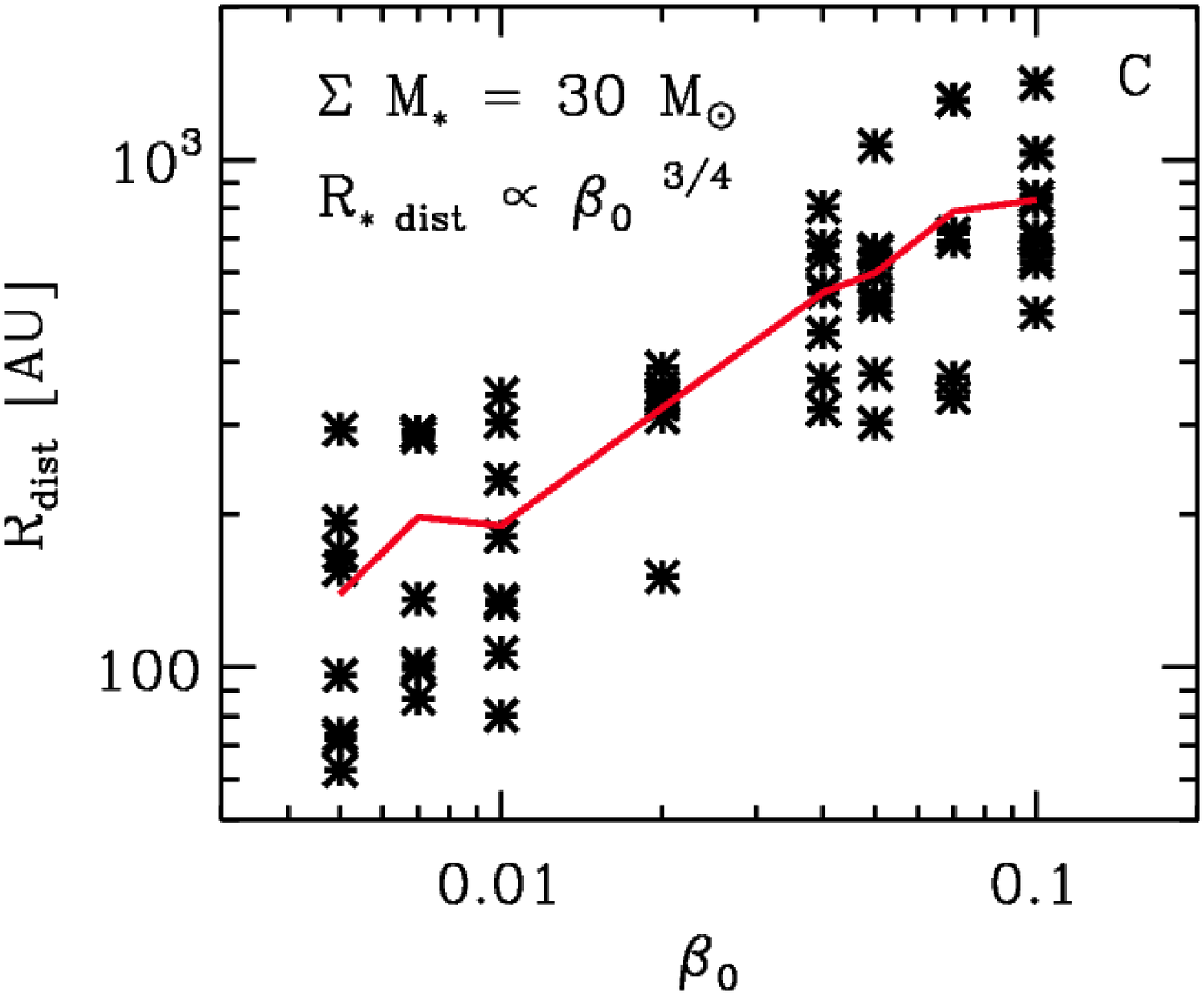}
           }
\caption{\label{sinknum} The fragmentation behavior is plotted for 
different degrees of initial rotation at the epoch when stellar system 
accretes $\sim$ 30 $M_\odot$. The number of sinks (A), time taken to 
accrete $\sim$ 30 $M_\odot$ (B) and the position of the sinks 
($R_{\rm dist}$) from the center of mass is shown in (C) as a 
function of the rotation parameter $\beta_0$. 
} 
\end{figure}

Figure~\ref{sinknum}A shows the number of sinks for different values 
of $\beta_0$, with rotationally supported clouds fragmenting the most. 
This is consistent with the recent study by \cite{bgsh15}, who have 
used {\em Arepo} simulations to investigate the dependence of the 
high accretion rate and efficient cooling of the gas on the 
fragmentation of the disk. However, Fig.~\ref{sinknum}A shows 
some scatter in the numbers, indicating the presence of statistical 
fluctuations that can be removed by pursuing more realizations to 
achieve a desired degree of accuracy. Figure~\ref{sinknum}B shows 
the time taken for all the clouds to accumulate $\sim$ 30 $M_{\odot}$. 
As expected, higher rotating clouds take longer  compared
to their slowly rotating counterparts. Figure~\ref{sinknum}C shows 
the distance at which the primordial protostars form from the center 
of the cloud. The red line represents the mean distances of all 
protostars ($R_{\rm dist}$), which follows a power-law relationship 
with the cloud's initial rotation, $R_{\rm dist} \propto \beta_0^{3/4}$. 
The protostars of the slowly rotating clouds form near the center 
($\leq$ 300 AU), while the others spread over larger distances of 
5000 AU, as the conservation of angular momentum acts to move 
protostars to larger radii. 

\begin{figure}
\centerline{
\includegraphics[width=3.6in]{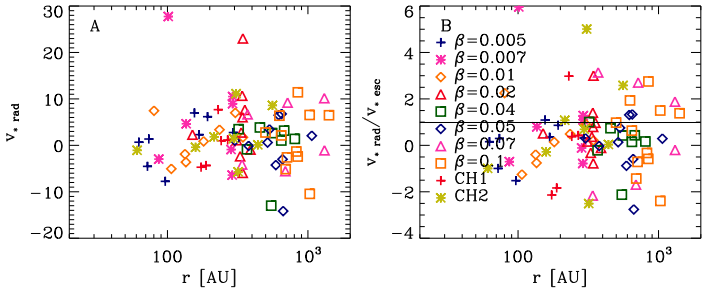}
           }
\caption{\label{sink_vesc} Radial velocity (A) and the ratio of the 
radial velocity to the escape velocity (B) of sink particles for 
different values of $\beta_0$ are plotted as a function of radius, 
at the epoch when the stellar system has accreted $\approx$ 30 
$M_\odot$. Some protostars with higher $\beta_0$ move away from the 
cluster, with the radial velocity exceeding the escape velocity. }
\end{figure}

Another trend we find in our simulations is that a number of 
protostars are ejected from the central gas cloud, as seen in 
cosmological simulations \citep{gswgcskb11}. Although the resolution 
used in those studies was higher than our resolution, we still 
find ejection from the cluster. Figure~\ref{sink_vesc} shows the 
radial velocity and the ratio of the radial velocity to the escape 
velocity for all protostars. The position of the sinks is measured 
from the center of mass of all the sinks. The escape velocity of 
the sinks is defined as 
$v_{\rm esc} = \sqrt{2GM_{\rm enc}(r)/r}$, where $M_{\rm enc}(r)$ 
is the total mass (gas + sinks) that is enclosed within the radius $r$. 
The radial velocities of the sinks formed with lower $\beta_0$ are 
below that required to be kicked out of the cluster. They tend to 
remain within the cluster and continue to accrete. For faster-rotating 
clouds, some protostars move from the cluster with radial velocities 
exceeding the escape velocity. There is, therefore, a greater chance that 
some protostars will be ejected, opening up the possibility that 
they could survive until the present day.

At this point we would like to point out that we have not mentioned 
the behavior of sinks for the $\beta_0=$ 0.2 case. The thermodynamical 
evolution of this cloud is considerably different from other cases, so 
it is important to compare the fragmentation behavior of $\beta_0=$ 0.2 
cloud with others. However, because of very fast rotation it has been 
extremely tough to run the simulation up to the point where the total 
mass of all sinks reaches $\sim$ 30 $M_{\odot}$. The simulation with 
$\beta_0=$ 0.2 stops much before compared to other simulations. For 
instance, simulation with $\beta_0=$ 0.2 stops when the total mass of 
sinks is only around 3$\hbox{--}$4 $M_{\odot}$  (and to reach up to that 
epoch of time; it takes around one month to run on a supercomputer
that is based on graphical processing units$\hbox{,}$ 
HPC$\hbox{--}$GPU Cluster Kolob). This is usually expected as 
the fast rotation can impede the collapse. We find a limiting value of  
$\beta_0=$ 0.1 for the simulation to run up to the epoch when the mass 
of sinks attains a value of 30 $M_{\odot}$. 
However, Figs. (4$\hbox{--}$7) still allow us to extrapolate and
predict the expected results.

\section{Summary and discussion}
\label{sec:summary}

We have minutely investigated  the dependence of the thermodynamical 
evolution of primordial star-forming gas on the initial degree of 
rotation of the cloud, and analyzed its influence on the resulting 
fragmentation of the circumstellar accretion disk. For this purpose, 
we  performed a set of three-dimensional hydrodynamical simulations 
of Pop~III gas collapse to pursue a systematic parameter study, 
specified by $\beta_0$, which  spans two orders of magnitude for the 
amount of initial rotation, including two simulations runs with  
realistic cosmological initial conditions.

The cloud's 
initial strength of rotation introduces a significant impact on 
the intricate combinations of the heating and cooling process, leading 
to a scatter in the temperature evolution of the collapsing gas. Clouds 
with slower rotation collapse faster and  get heated as a result of the 
compressional heating. We also find that the dynamical evolution of the 
gas strongly depends on the initial strength rotation. Clouds with higher 
rotation form a Keplerian disk that becomes gravitationally unstable by 
accreting the infalling mass. Therefore, any change in the thermodynamical 
evolution introduces substantial difference in the number of Jeans mass, 
which determines the susceptibility to the fragmentation of the gas 
between the clouds with highest and lowest initial rotation.  

In summary, a higher degree of rotation can hinder the infall, 
lead to a cooler gas, and  result in more fragmentation.  
In addition, we find that the protostars with higher rotational 
support have larger spiral arms with lower accretion rates. We 
also point out that the newborn protostars are distributed in 
such a way to conserve the angular momentum, and some of them 
could have survived until today if they were of sufficiently low mass.

We conclude that the initial conditions of the primordial gas in 
the minihalos should be chosen scrupulously so as to simulate 
the long-term evolution and final fate of the primordial stars.

Despite considerable computational efforts involved, we emphasize 
 that we cannot accurately predict the final mass of the 
primordial protostars.   
We have neglected the effect of the magnetic fields, which can 
be important in minihalos \citep{mmi08,sur10,sbsakbs10}. Recent 
cosmological simulations show the importance of the amplification 
of even small seed fields \citep{fssbk11,ssfgkb12}. In addition, 
the radiative feedback can significantly affect the thermal as 
well as chemical evolution of the gas \citep{wa08,whm10,hoyy11},
which is not included in our simulations. Notwithstanding, 
our approach to the problem enables us to provide good estimates 
for the overall trend of the accretion rate, thermodynamical 
evolution, and fragmentation behavior of  gas in the rotating clouds 
in which we are particularly interested. Recent radiation 
hydrodynamic simulations \citep{hirano13,greif14} demonstrated 
the effect of UV radiative feedback on the mass accretion, thus 
constraining the mass spectrum of the first stars \citep{susa13}. 
Moreover, there is still a discrepancy regarding the three-body 
$\rm H_2$ rate coefficient \citep{tcggakb11,bsg14,dnck15}. However, 
the recent study by \citet{forrey13} provides the currently best 
available rate that is in good agreement with \citet{pss83} at high 
temperatures and \citet{abn02} at lower temperatures. It is 
therefore of strategic interest to accurately simulate the formation 
of the first stars in the Universe with the best available rates, 
inclusion of the magnetic fields, and radiative feedback in the 
next-generation avant-garde SPH codes.

\smallskip
The author wishes to thank Prateek Sharma, Biman Nath, David Sobral, 
and Dominik Schleicher for thoroughly checking the manuscript and 
for enormous constructive suggestions. The author also acknowledges 
Kazu Omukai, Athena Stacy, 
and the referee for helpful and worthwhile comments. The present 
work is supported by the Indian Space Research Organization (ISRO) grant 
(No.~ISRO/RES/2/367/10-11) and Department of Science and Technology (DST) 
grant (Sr/S2/HEP-048/2012). The author is grateful to the Centre for 
Theoretical Studies (CTS) at the Indian Institutes of Technology 
$\hbox{--}$ Kharagpur and Raman Research Institute for the financial 
support and hospitality. The author would also like to thank the 
Department of Physics, Indian Institute of Technology (Banaras Hindu 
University) at Varanasi and the Inter-University Center for Astronomy 
and Astrophysics (IUCAA) at Pune for  local hospitality.

\footnotesize{

}

\end{document}